\documentclass[a4paper,fleqn]{cas-sc}
\usepackage[round]{natbib}
\usepackage{bm}

\newcommand\ket [1] {|#1 \rangle }
\newcommand\bra [1] {\langle #1 |}

\newcommand{\bb}[1]{\mathbf{#1}}

\shorttitle{Recent Developments in Fractional Chern Insulators}    

\shortauthors{Z. Liu and E. J. Bergholtz}  

\begin{document}
\let\WriteBookmarks\relax
\def\floatpagepagefraction{1}
\def\textpagefraction{.001}

\title[mode = title]{Recent Developments in Fractional Chern Insulators}

%\tnotemark[1,2]

%\tnotetext[1]{This document is the results of the research project
   % funded by the National Science Foundation.}

  %\tnotetext[2]{The second title footnote which is a longer text
    %matter to fill through the whole text width and overflow into
    %another line in the footnotes area of the first page.}

  \author[1]{Zhao Liu}[orcid=0000-0002-3947-4882]
  \cormark[1] 
  %\fnmark[1] 
  \ead{zhaol@zju.edu.cn} 
  %\ead[url]{www.cvr.cc,www.tug.org.in}

  %\credit{Conceptualization of this study, Methodology, Software}

  \address[1]{Zhejiang Institute of Modern Physics, Zhejiang University, Hangzhou 310027, China}

  \author[2]{Emil J. Bergholtz}[orcid=0000-0002-9739-2930]
  \ead{emil.bergholtz@fysik.su.se}

  \address[2]{Department of Physics, Stockholm University, AlbaNova University Center, 106 91 Stockholm, Sweden}

  \cortext[cor1]{Corresponding author} 
  %\cortext[cor2]{Principal corresponding author} 
  %\fntext[fn1]{This is the first author footnote. 
    %but is common to third author as well.}

  %\fntext[fn2]{Another author footnote, this is a very long footnote
    %and it should be a really long footnote. But this footnote is not
    %yet sufficiently long enough to make two lines of footnote text.}

  %\fntext[fn3]{K. Berry is the editor of \TeX Live.}

  %\nonumnote{This note has no numbers. I}

 \begin{abstract}
Fractional Chern insulators (FCIs) are lattice generalizations of the conventional fractional quantum Hall effect (FQHE) in two-dimensional (2D) electron gases. They typically arise in a 2D lattice without time-reversal symmetry when a nearly flat Bloch band with nonzero Chern number is partially occupied by strongly interacting particles. Band topology and interactions endow FCIs exotic topological orders which are characterized by the precisely quantized Hall conductance, robust ground-state degeneracy on high-genus manifolds, and fractionalized quasiparticles. Since in principle FCIs can exist at zero magnetic field and be protected by a large energy gap, they provide a potentially experimentally more accessible avenue for observing and harnessing FQHE phenomena. Moreover, the interplay between FCIs and lattice-specific effects that do not exist in the conventional continuum FQHE poses new theoretical challenges. In this chapter, we provide a general introduction of the theoretical model and numerical simulation of FCIs, then pay special attention on the recent development of this field in moir\'e materials while also commenting on potential alternative implementations in cold atom systems. With a plethora of exciting theoretical and experimental progress, topological flat bands in moir\'e materials such as magic-angle twisted bilayer graphene on hexagonal boron nitride have indeed turned out to be a remarkably versatile platform for FCIs featuring an intriguing interplay between topology, geometry, and interactions.
  
  \end{abstract}

 \maketitle
%\linenumbers

\section{Introduction}
Restricting electrons in a two-dimensional (2D) plane penetrated by a uniform magnetic field and cooling the system to low temperature seems like a simple setup. However, it provides one of the most significant discoveries in condensed matter physics over the past half century. Remarkably, the Hall resistance $R_H$ of this 2D electron gas (2DEG) has a novel dependence on the magnetic field $B$. Instead of increasing linearly with $B$ as predicted by the classical theory, $R_H$ develops multiple plateaus near proper rational filling $\nu$, on which its value is precisely quantized at $R_H= h/(\nu e^2)$. Here $h$ is the Planck constant, $-e$ is the electron charge, and $\nu=N/N_\phi$ with $N$ and $N_\phi$ the number of electrons and magnetic flux quanta, respectively. This phenomenon is known as the quantum Hall effect (QHE). The plateaus near integer $\nu=n$, called the integer QHE (IQHE), were first discovered in 1980 by K. von Klitzing~\cite{IQHE}. These IQHE states can be understood in a single-particle picture: they result from the Landau quantization of an electron in the magnetic field, and correspond to $n$ fully filled Landau levels (LLs). On the boundary, these $n$ LLs come with $n$ chiral edge modes. The precisely quantized $R_H$ originates from a topological invariant -- the Chern number $\mathcal{C}$ which is one for each LL~\cite{Thouless,Avron1983,QianNiu1985}. This Chern number cannot change so long as the energy gap in the bulk, which is the LL spacing in this case, does not close. This gives rise to the robustness of the IQHE plateaus. In this sense, the IQHE states are the first observed topological band insulators. By contrast, the QHE near fractional $\nu=f$, observed by Tsui and St\"ormer in 1983~\cite{FQHE} and dubbed the fractional quantum Hall effect (FQHE), has a very different origin. As the single-particle states in each LL are exactly degenerate, it must be the electron-electron interaction that lifts the huge degeneracy in the partially filled LL and stabilize the FQHE. While this strongly correlated nature makes the FQHE notoriously difficult to study, it also endows the FQHE an exotic order: the intrinsic topological order~\cite{Wen1990}, as characterized by non-trivial ground-state degeneracies depending on the genus of the manifold on which the FQH states reside~\cite{Topologicaldeg}, long-range entanglement~\cite{TEEKitaev,TEELevin}, and fractionalized quasiparticle excitations~\cite{Laughlin83,MOORE1991362,Feldman_2021}. Motivated by understanding and utilizing exotic topological orders, the FQHE has grown to become one of the arguably most exuberant fields of modern condensed matter physics. 

With more FQHE plateaus discovered in 2DEGs and deeper understanding obtained about the nature of observed states, relaxing the limitations of the strong magnetic field and ultralow temperature required to realize the FQHE remains a long-standing dream. In conventional 2DEG setups, $B\sim 15 \ {\rm T}$ and $T\lesssim 1 \ {\rm K}$ are necessary for the robust $\nu=1/3$ plateau~\cite{FQHE}. We need stronger magnetic fields to reach smaller fillings, and temperature of at least one order of magnitude lower and cleaner samples to see more fragile FQH states. Hence, a natural question arises: can we realize the FQHE at zero (at least much weaker) magnetic fields and higher temperature? Intuitively the solution to this question calls for a platform change. The first important step in this direction was provided by Haldane in his seminal work in 1988, where he constructed a simple tight-binding model on the honeycomb lattice and showed that the IQHE can exist in the lattice at zero net magnetic field~\cite{Haldane1988}. In this model, the time-reversal symmetry (TRS) breaking caused by the magnetic field in the conventional IQHE is achieved by complex hopping which can be emulated by spin-orbit coupling in the lattice, and isolated Bloch bands carrying $\mathcal{C}=\pm 1$ play the role of LLs. Such lattice versions of the conventional IQHE are now called Chern insulators, which were first reported in 2013 in thin films of chromium-doped ${\rm (Bi,Sb)}_2{\rm Te}_3$~\cite{CI_Chang} and later also in cold-atom systems~\cite{Jotzu2014}.

Haldane's construction of Chern insulators raises the possibility of obtaining the FQHE in 2D lattice systems with broken TRS, dubbed fractional Chern insulators (FCIs), down to zero magnetic field. Of course, a lattice model should satisfy several conditions in order to host FCIs. The most natural requirement is a nearly flat Bloch band with nonzero Chern number, i.e., a nearly flat topological (Chern) band which can mimic an exactly flat LL in a 2DEG. The weak band dispersion enhances the interaction effect, so that intuitively interacting particles have a chance to form FQHE-like states when they partially occupy the band. Despite of important early work on FCIs, including the construction of a parent Hamiltonian in lattices containing only short range interactions and an exponentially decaying tail of hopping terms~\cite{KapitMueller}, the research in this direction became mainstream only in 2011 when it was emphasized that nearly flat Chern bands can exist in tight-binding models with strictly short-ranged hopping~\cite{flatband_Wen,flatband_Sun,flatband_Titus}. Subsequent numerical works indeed confirmed the existence of FCIs in various $|\mathcal{C}|=1$ flat bands as lattice analogs of conventional FQH states, including the Laughlin state, Moore-Read state, and hierarchy-composite-fermion series~\cite{Sheng2011,FCIPRX,FCIChern_Wang,FCIChern_Wang2,FCIZoology,Andreas2013,FCIJain_Liu}. The adiabatic continuity between these FCIs and the corresponding conventional FQH states~\cite{FCIWannier_Gunnar,Wannier_Wu,Continuity_Zhao,FCIHighC_YangLe} was established via mapping single-particle states in a Chern band to those in a LL~\cite{Wannier_Qi,Wannier_Wu,Wannier_Qi2,FCIHighC_YangLe,HaldaneStatistics}. Crucially, the energy gap protecting these FCIs may be greatly increased compared to conventional FQH states in 2DEGs, as it is determined by Coulomb interaction on the lattice scale rather than the magnetic length in 2DEGs which is typically two orders of magnitude larger. In fact, a naive estimate does not rule out FCIs at room-temperature~\cite{flatband_Wen}. 

The benefit of studying FCIs soon exceeds just finding lattice analogs of the already known conventional FQH states. The remarkable distinctions between a Chern band and a LL raise possibilities of new physics and meanwhile pose new challenges. Unlike for LLs, the quantum geometry (Berry curvature, Fubini-Study metric, etc.) of a Chern band is in general fluctuating in the Brillouin zone. Shortly after the first numerical identifications of FCIs, the key role played by this inhomogeneous quantum geometry on the stability of FCIs was noticed~\cite{Roy2012,Roy2014,Jackson2015}. Geometry might sound less important than topology, because under deformation of the band it is not invariant like Chern number even if the band keeps isolated. However, analytical and numerical analysis has unveiled it as a crucial criterion to judge how good a Chern band is for hosting FCIs. Thanks to the studies on FCIs, now we are understanding more about the effect of fluctuating band geometry on the many-body physics~\cite{BandGeometry_Lee,JieBandGeometry,FieldTunedFCI,EmilBandGeometry,BandGeometry_Ahmed}, which is absent in the conventional FQHE due to the uniform geometry of a LL. Apart from the non-uniform quantum geometry, a Bloch band can support high Chern number $|\mathcal{C}|>1$~\cite{WangFa2011,Trescher2012,Yang2012}, while each LL can only carry $|\mathcal{C}|=1$. This lattice-specific feature implies qualitatively new FCI states beyond the FQH paradigm. Excitingly, plethora of such states have been confirmed in high Chern number bands~\cite{FCIHighC_Zhao,YiFei2012,FCIHighC_Sterdyniak,FCIHighC_YangLe,GunnarHof,Wu2015,FCIHighC_Emil,ModelFCI_Zhao}, but the complete understanding of them still needs more effort. Moreover, lattice defects in these high-$\mathcal{C}$ FCIs may act as worm-hole-like objects, suitably dubbed ``genons,'' that effectively change the genus of space and obey non-Abelian exchange statistics~\cite{PhysRevX.2.031013,PhysRevB.87.045130,Genon_Zhao,Gappedboundary_Zhao,PhysRevB.99.081114}. This could lead to a new scheme of topological quantum computation.

The field of FCI has flourished over the last decade, motivated by its potential to manifest higher-temperature FQHE-like phenomena at much weaker magnetic fields and novel physics beyond the conventional FQHE. A main trend in the development of this field is to gather inspiration from other research fields, like material science and ultracold atoms, such that we can find suitable lattice platforms to really produce FCIs in experiments. Remarkably, exciting progress has been made along this direction in recent years. The first experimental observation of FCI states was reported in 2018 for Bernal-stacked bilayer graphene, which, however, still needs a strong magnetic field of $B\sim 30 \ {\rm T}$ to create the required topological flat bands~\cite{FCI_Eric}. The rapid development of techniques to fabricate van-der-Waals heterostructures with moir\'e patterns~\cite{Geim2013,Novoselov2016} provides new materials that can harbor intrinsic flat Chern bands at zero magnetic field~\cite{TBGCBand_Senthil}, in which recent numerical works have identified the existence of zero-field FCIs~\cite{FCITBG_Ahmed,TBGFCI_Ceceil,FCITDBG_Zhao,FieldTunedFCI}. Consistent with earlier theoretical predictions~\cite{FCITBG_Ahmed,TBGFCI_Ceceil,ChiralTBG_Ashvin2}, multiple FCI states were indeed experimentally observed in 2021 with twisted bilayer graphene (TBG)~\cite{Bistritzer2011,TBGReview_MacDonald} aligned with hexagonal boron nitride at weak magnetic fields down to $5 \ {\rm T}$~\cite{FCIexp_Xie}, in which the weak magnetic field was used only to improve the band geometry such that FCIs became favored. This experiment paves the way toward real zero-field FCIs, and meanwhile reports some exotic FCIs which need further investigation. Currently, the research on a broad range of moir\'e materials and the strongly-correlated phases thereof is making FCI an active field. 

This chapter will provide a general introduction to the basic concepts, the methodology and the experimental observation of FCIs, mainly responding to the following questions: (i) what is the appropriate theoretical microscopic model to study FCIs? (ii) what conditions should this model satisfy to favor FCIs? (iii) how to identify FCIs and their competing phases in numerical simulations and experiments? (iv) how different FCIs are from the conventional FQHE? Meanwhile we will focus on the very recent development through our introduction. It is impossible to cover everything in this field within a chapter. For topics that are not discussed in detail, we recommend the readers to refer to earlier review articles~\cite{PARAMESWARAN2013816,zhao_review,Neupertreview_2015} and references therein. 

\section{Microscopic model}
Here we present the general microscopic model for studying FCIs in 2D lattices with broken TRS. In most cases, we need to get a many-body Hamiltonian projected to a partially filled Bloch band to capture the low-energy physics within the band, in analogy to the LL projection in 2DEGs. There are two typical ways to achieve this in the literature, which we will describe below. 

\subsection{Tight-binding formalism} 
Tight-binding formalism is a common method to model particles in lattice systems. It is practical especially when the underlying lattice is not too complicated (i.e., only a few sites in each unit cell), and is highly relevant to cold atoms in optical lattices. This formalism has been used to study FCIs in various toy lattice models, like the kagome, checkerboard, and honeycomb lattice~\cite{Sheng2011,FCIPRX,FCIChern_Wang,FCIChern_Wang2,FCIZoology,Andreas2013}. Under the tight-binding scenario, it is natural to consider a many-body Hamiltonian of the form
\begin{eqnarray}
H=\sum_{n,m} \sum_{\alpha,\beta}  t^{\alpha\beta}_{nm}c^\dagger_{n,\alpha}c_{m,\beta}+ \sum_{n,m}\sum_{\alpha,\beta}V^{\alpha\beta}_{nm}c^\dagger_{n,\alpha}c^\dagger_{m,\beta}c_{m,\beta}c_{n,\alpha},
\label{tH}
\end{eqnarray}
where $n,m=1,\cdots,N_u$ label the unit cells of the lattice at positions ${\bf R}_n$ and ${\bf R}_m$, $\alpha,\beta=1,\cdots,N_b$ label the internal states (for example, inequivalent sites) in each unit cell, and $c^\dagger_{n,\alpha}$ ($c_{n,\alpha}$) creates (annihilates) a particle in unit cell ${\bf R}_n$ and state $\alpha$. In Eq.~(\ref{tH}), the first term is the single-particle Hamiltonian describing the hopping of particles in the lattice, and the second term is the two-body Hubbard-like interaction between particles. Translation invariance of both the hopping and the interaction is usually assumed, that is, $ t^{\alpha\beta}_{nm}$ and $V^{\alpha\beta}_{nm}$ only depend on ${\bf R}_n-{\bf R}_m$. Then we can switch to the reciprocal space by the Fourier transform $c^\dagger_{{\bb k},\alpha}=\frac 1 {\sqrt{N_u}}\sum_n e^{i{\bb k}\cdot {\bb R}_n}c^\dagger_{n,\alpha}$, which creates a particle in state $\alpha$ with crystal momentum ${\bf k}=(k_x,k_y)$ restricted in the first Brillouin zone (1BZ) of the lattice. This reexpresses Eq.~(\ref{tH}) as
\begin{eqnarray}
H=\sum_{{\bb k}\in {\rm 1BZ}}\sum_{\alpha,\beta} \mathcal H_0^{\alpha\beta}({\bf k})c^\dagger_{{\bb k},\alpha}c_{{\bb k},\beta}
+\sum_{\{ {\bf k}_i\}\in{\rm 1BZ}}\sum_{ \alpha,\beta}V_{\bb k_1\bb k_2\bb k_3\bb k_4}^{\alpha\beta}c^\dagger_{{\bb k_1},\alpha}c^\dagger_{{\bb k_2},\beta}c_{{\bb k_3},\beta}c_{{\bb k_4},\alpha}. 
\label{tHk}
\end{eqnarray}
In Eq.~({\ref{tHk}), the single-particle and interaction matrix elements are 
\begin{eqnarray}
\mathcal{H}_0^{\alpha\beta}({\bf k})=\sum_{n} t^{\alpha\beta}_{n0}e^{-i{\bb k}\cdot {\bb R}_n}
\label{H0}
\end{eqnarray}
and
\begin{eqnarray}
V_{\bb k_1\bb k_2\bb k_3\bb k_4}^{\alpha\beta}=\frac{1}{N_u}\delta_{\bb k_1+\bb k_2,\bb k_3+\bb k_4}' \sum_{n} V^{\alpha\beta}_{n0}e^{-i(\bb k_1-\bb k_4)\cdot \bb R_n},
\label{V}
\end{eqnarray}
respectively, where we have set ${\bf R}_m={\bf 0}$ and $\delta'_{{\bf k},{\bf k}'}$ is the 2D periodic Kronecker delta function with period of primitive reciprocal lattice vectors. 

The single-particle band structure of the system can be extracted by diagonalizing the $N_b\times N_b$ matrix $\mathcal H_0^{\alpha\beta}({\bf k})$ for each $\bb k \in{\rm 1BZ}$ separately. The resulting eigenvalues $\{\epsilon_s({\bf k})\}$ are just the band energies, and the eigenvector $\psi_s({\bf k})$ gives the Bloch state of the $s$-th band as $|\psi_s({\bf k})\rangle=\sum_{\alpha}\psi_s^\alpha({\bf k}) c_{\bf k,\alpha}^\dagger |{\rm vacuum}\rangle$. With the band eigenvectors at hand, one can define a new set of operators $\gamma_{{\bf k},s}^\dagger$ that create a particle with crystal momentum ${\bf k}$ in a specific band $s$, such that
\begin{eqnarray} 
\label{ctogamma}
\gamma_{{\bf k},s}^\dagger=\sum_\alpha \psi_s^{\alpha}({\bf k})c_{{\bf k},\alpha}^\dagger, \ 
c_{{\bf k},\alpha}^\dagger=\sum_s [\psi_s^{\alpha}({\bf k})]^* \gamma_{{\bf k},s}^\dagger.
\end{eqnarray} 
In terms of $\gamma_{{\bf k},s}^\dagger$ and $\gamma_{{\bf k},s}$, one finds the many-body Hamiltonian Eq.~({\ref{tHk}) is rewritten as
\begin{eqnarray}
H=\sum_{{\bb k}\in {\rm 1BZ}}\sum_{s} \epsilon_s({\bf k})\gamma^\dagger_{{\bb k},s}\gamma_{{\bb k},s}
+\sum_{\{ {\bf k}_i\}\in{\rm 1BZ}}\sum_{\{s_i\}}\sum_{ \alpha,\beta}V_{\bb k_1\bb k_2\bb k_3\bb k_4}^{\alpha\beta}
[\psi_{s_1}^{\alpha}({\bf k}_1)]^* [\psi_{s_2}^{\beta}({\bf k}_2)]^* \psi_{s_3}^{\beta}({\bf k}_3) \psi_{s_4}^{\alpha}({\bf k}_4)
\gamma^\dagger_{{\bb k_1},s_1}\gamma^\dagger_{{\bb k_2},s_2}\gamma_{{\bb k_3},s_3}\gamma_{{\bb k_4},s_4}. 
\label{H2}
\end{eqnarray}

The transform of the many-body Hamiltonian from Eq.~(\ref{tH}) to Eq.~(\ref{H2}) makes it convenient to deal with a situation that we are primarily interested in, that is, a single isolated band is partially filled by interacting particles while other bands are either fully occupied or empty. In this case, it makes sense to project the whole Hamiltonian to the partially filled band -- this is where all the low-energy actions are expected to take place. Naively one may think this is valid only when the interaction strength is much smaller than the band gap such that band mixing is absent. However, the band projection actually works far beyond this limit as particles often prefer to restrict the low-energy actions in the partially filled band even if the interaction strength exceeds the band gap~\cite{PhysRevLett.112.126806}, similar to the situation in the conventional FQHE. 

To do the band projection, we restrict the band indices in Eq.~(\ref{H2}) to the partially filled band, reaching the projected Hamiltonian 
\begin{eqnarray}
H=\sum_{{\bb k}\in {\rm 1BZ}} \epsilon({\bf k})\gamma^\dagger_{{\bb k}}\gamma_{{\bb k}}
+\sum_{\{ {\bf k}_i\}\in{\rm 1BZ}}\bar{V}_{\bb k_1\bb k_2\bb k_3\bb k_4}
\gamma^\dagger_{{\bb k_1}}\gamma^\dagger_{{\bb k_2}}\gamma_{{\bb k_3}}\gamma_{{\bb k_4}},
\label{projH}
\end{eqnarray}
where we have dropped the band index for simplicity and 
\begin{eqnarray}
\bar{V}_{\bb k_1\bb k_2\bb k_3\bb k_4}=\sum_{\alpha,\beta}V_{\bb k_1\bb k_2\bb k_3\bb k_4}^{\alpha\beta}
[\psi^{\alpha}({\bf k}_1)]^* [\psi^{\beta}({\bf k}_2)]^* \psi^{\beta}({\bf k}_3) \psi^{\alpha}({\bf k}_4).
\label{projV}
\end{eqnarray}
Eq.~(\ref{projH}) is the starting point in most literatures to investigate FCIs and other correlated phases occurring in lattice systems. In the flat-band limit where the band dispersion is sufficiently weak, we can even only keep the interaction term in Eq.~(\ref{projH}). In this case, the band-projected Hamiltonian has almost the same form as the LL-projected Hamiltonian of the conventional FQHE, for which ${\bf k}$ is replaced by the orbital index of the LL. 

Although majority of reported FCI states is captured by Eq.~(\ref{projH}), exceptions beyond the single-band description exist. First, projection to multiple bands is necessary if an isolated set of bands is partially occupied, leading to multicomponent FCI states~\cite{FCIZ2_Regnault,FCIHalperin_Zeng,FCIHalperin_Zeng2,FCIHalperin_Zeng3,FCIHalperin_Zeng4,FCIHalperin_Zeng5}. The generalization of Eq.~(\ref{projH}) to this case is straightforward. Second, sometimes strong interactions may dramatically modify the non-interacting band structure and produce an emergent Chern band in which FCIs can develop, even if the non-interacting band has zero Chern number or even not isolated~\cite{FCIPES_Simon,FCIPumping_Zhu}. We cannot rely on the single-band projection to study these states. For simplicity, we will not discuss these situations in this chapter. 

\subsection{Effective continuum model}
Unlike in toy lattice models, realistic solid-state materials are often too complicated to be formulated as a simple tight-binding Hamiltonian. This is the case in moir\'e materials where a unit cell contains a few hundreds sites. For moir\'e systems, theorists usually rely on first-principle calculations to construct an effective continuum single-particle Hamiltonian $H_0$ in the reciprocal space to describe the low-energy bands (for instance, see Refs.~\cite{Bistritzer2011,TBGCBand_Senthil,TBGCBand_Senthil2,AshvinTDBG,TDBG_Haddadi,TDBG_Jung}). We will give a concrete example of $H_0$ for twisted bilayer graphene in Sec.~\ref{section:tbghbn}, but here we give some general statements. 

Like in the tight-binding formalism, the effective $H_0$ is also expressed in terms of the particle's creation and annihilation operators -- $c^\dagger_{{\bb k},\alpha}$ and $c_{{\bb k},\alpha}$, where $\alpha$ labels internal degrees of freedom of a particle. Similar to Eq.~(\ref{ctogamma}), $c^\dagger_{{\bb k},\alpha}$ and $c_{{\bb k},\alpha}$ can in turn be related to the band operators $\gamma^\dagger_{{\bb k},s}$ and $\gamma_{{\bb k},s}$ after diagonalizing $H_0$. Regarding the two-body interaction in continuum, we can write it by the standard second-quantization formula as 
\begin{eqnarray}
H_{\rm int}=\frac{1}{2}\sum_{{\bf q}\in R^2} \tilde{V}({\bf q}):\rho({\bf q})\rho(-{\bf q}):,
\label{Hint0}
\end{eqnarray}
where the summation of ${\bf q}$ is in the whole reciprocal space $R^2$ instead of only the 1BZ, $\tilde{V}({\bf q})=\frac{1}{A}\int V({\bf r})e^{-i{\bf q}\cdot{\bf r}}d{\bf r}$ is the Fourier transform of the interaction potential $V({\bf r})$ with $A$ the area of the 2D system, $\rho({\bf q})=\sum_\alpha c^\dagger_{{\bb k}+{\bf q},\alpha}c_{{\bb k},\alpha}$ is the density operator, and $::$ is the normal ordering. Then we can project the many-body Hamiltonian $H=H_0+H_{\rm int}$ to the partially filled band by the relation between $c^\dagger_{{\bb k},\alpha},c_{{\bb k},\alpha}$ and $\gamma^\dagger_{{\bb k},s},\gamma_{{\bb k},s}$ and restricting the band index $s$ to the partially filled band. The obtained band-projected Hamiltonian has exactly the same form as Eq.~(\ref{projH}). 

\section{Conditions of the existence of FCIs}
\label{sec:conditions}
Given the Hamiltonian Eq.~(\ref{projH}) of interacting particles partially occupying a single isolated band, whether FCIs are favored as the ground states strongly depends on the system details, including the filling factor, the shape of the finite lattice, and properties of the band and the interaction. Insight of proper filling factors can be obtained in analogy to the conventional FQHE. For a finite lattice, we expect that its shape should be close to the 2D isotropic limit.  In the following, we will summarize the known conditions which the band and the interaction should satisfy to harbor FCIs. 

\subsection{Requirements for the band}
\label{sec:bandconditions}
For a partially filled isolated band $s$, numerical works and analytical studies have identified three of its properties -- bandwidth, topology, and quantum geometry, that are highly relevant to the existence of FCIs. The bandwidth $W$ is determined by the band energies via $W=\max_{\bb{k}, \bb{k}' \in \rm{1BZ}}[\epsilon_{s}(\bb{k})-\epsilon_{s}(\bb{k}')]$. By contrast, the topology and quantum geometry are encoded in band eigenstates~\cite{Ran2013,Roy2014,Jackson2015,PhysRevB.104.045103}. Denoting the cell-periodic part of $|\psi_s({\bf k})\rangle$ as $\ket {u_s({\bf k})}$, the quantum geometric tensor of band $s$ is defined as
\begin{eqnarray} 
\mathcal{Q}_s^{ab}({\bf k}) =\langle \partial_{{\bf k}}^a u_s({\bf k})|\partial_{{\bf k}}^b u_s({\bf k})\rangle 
-\bra {\partial_{{\bf k}}^a u_s({\bf k})} u_s({\bf k})\rangle \bra{u_s({\bf k})}\partial_{{\bf k}}^b u_s({\bf k})\rangle
\equiv g_s^{ab}({\bf k})-\frac{i}{2}\Omega_s^{ab}({\bf k})
\end{eqnarray}
with $a,b=x,y$. Up to a constant factor $-1/2$, the imaginary part of $\mathcal{Q}_s^{ab}({\bf k})$ is just the Berry curvature 
\begin{eqnarray} 
\Omega_s^{ab}({\bf k})=i\left[\langle \partial_{{\bf k}}^a u_s({\bf k})|\partial_{{\bf k}}^b u_s({\bf k})\rangle
-\langle \partial_{{\bf k}}^b u_s({\bf k})|\partial_{{\bf k}}^a u_s({\bf k})\rangle\right]
\label{berry}
\end{eqnarray} 
which resembles the effect of magnetic field in reciprocal space. The band topology is then characterized by the integral of Berry curvature over the 1BZ, called Chern number
\begin{eqnarray} 
\mathcal{C}_s=\frac{1}{2\pi}\int_{\rm 1BZ} \Omega_s^{xy}({\bf k}) dk_x dk_y,
\label{chern}
\end{eqnarray} 
which must be an integer. Chern number is a topological invariant in the sense that it cannot be changed by modifying the band dispersion unless the band gap is closed. 
On the other hand, the real part of the quantum geometric tensor is the Fubini-Study (FS) metric
\begin{eqnarray} 
g^{ab}_s({\bf k})=\frac{1}{2}\big(\langle \partial_{{\bf k}}^a u_s({\bf k})|\partial_{{\bf k}}^b u_s({\bf k})\rangle 
-\bra {\partial_{{\bf k}}^a u_s({\bf k})} u_s({\bf k})\rangle \bra{u_s({\bf k})}\partial_{{\bf k}}^b u_s({\bf k})\rangle
+a\leftrightarrow b\big)
\label{fs}
\end{eqnarray} 
of the band. As $|\langle u_s({\bf k})| u_s({\bf k}+d{\bf k})\rangle|\approx 1-\frac{1}{2}g^{ab}_s({\bf k})dk_a dk_b$, the FS metric measures the distance between two $|u_s\rangle$ states at different ${\bf k}$ points. Moreover, the Berry curvature and the FS metric are related by two important inequalities~\cite{PARAMESWARAN2013816,Roy2014,Jackson2015,BandGeometry_Lee,JieBandGeometry}
\begin{eqnarray} 
{\rm tr}g^{ab}_s({\bf k})\ge |\Omega_s({\bf k})|, \ \det g^{ab}_s({\bf k})\ge \frac{1}{4}|\Omega_s({\bf k})|^2,
\label{fsb}
\end{eqnarray} 
the saturations of which are referred to as the trace condition and the determinant condition, respectively. In fact, the second inequality can be derived from the first one~\cite{Roy2014,JieBandGeometry}, thus the determinant condition is definitely satisfied when the trace condition is.  

\begin{table}[htp]
\caption{The comparison between a Landau level and a Bloch band from aspects of band dispersion, topology, quantum geometry, and PH asymmetry of the projected two-body interaction. Here we assume no disorder or boundary effects.}
\label{tbl1}
\begin{tabular*}{\tblwidth}{m{3cm}m{6cm}m{6cm}c}
\toprule
  &  Landau level  \centering &  Bloch band \centering &  \\ % Table header row
\midrule
Dispersion & No dispersion, i.e., all single-particle states within each LL is perfectly degenerate. & Finite bandwidth in general.  \centering &\\
\hline
Topology & Chern number $|\mathcal{C}|=1$ for each LL. The sign is determined by the direction of the magnetic field. & Chern number of a Bloch band can be an arbitrary integer. Apart from $\mathcal{C}=0$ and $|\mathcal{C}|=1$, there are also Bloch bands with higher Chern number $|\mathcal{C}|>1$.
&\\
\hline
Quantum geometry & Uniform (${\bf k}$-independent) nonzero Berry curvature and FS metric. In particular, both inequalities in Eq.~(\ref{fsb}) are saturated for the lowest LL. & In general, both the Berry curvature and FS metric fluctuate in the 1BZ. While the trace and determinant conditions in Eq.~(\ref{fsb}) are satisfied for all two-band models at any ${\bf k}$~\cite{BandGeometry_Lee,JieBandGeometry}, in general this does not hold for multiband models. \\
\hline
PH asymmetry of projected two-body interactions & No \centering & Yes \centering&\\
\bottomrule
\end{tabular*}
\end{table}

Considering the formation of the conventional FQHE in LLs, we naturally expect that the existence of FCIs builds on how well a Bloch band can reproduce the physics of a LL. In Table~\ref{tbl1}, we compare a Bloch band with a LL. Based on their differences in aspects of dispersion, topology and quantum geometry, the following conditions appear to be necessary for a Bloch band to mimic a LL and host FQHE-like phenomena.
\begin{itemize}
\item {\bf Bandwidth.} The bandwidth $W$ should be smaller than the interaction strength $V$, such that the interaction dominates the physics. Otherwise, the ground state is expected to be a usual Fermi liquid of electrons or pile of bosons in the band bottom. Given a typical interaction scale in the system, it is easier to satisfy $W\ll V$ if the band is tuned to be flatter. 
\end{itemize}

\begin{itemize}
\item {\bf Band topology}. Typically we require that the band should carry nonzero Chern number, although this is not a fundamental necessity if strong interactions can induce an emergent Chern band ~\cite{FCIPES_Simon,CDW_Stefanos4}. To get nonzero Chern number, we need to break the TRS, which can be achieved by complex hopping via spin-orbit coupling, ferromagnetism, artificial gauge fields, or interaction-induced spontaneous TRS breaking.  
\end{itemize}

\begin{itemize}
\item {\bf Band geometry}. The Berry curvature and FS metric should not strongly fluctuate in the 1BZ. Meanwhile, the trace and determinant conditions in Eq.~(\ref{fsb}) should be approximately satisfied if we need to mimic the lowest LL (LLL). The key role of band geometry on the stability of FCIs has been demonstrated in various lattice models, for instance, see Refs.~\cite{Jackson2015,CDW_Patrick,PhysRevB.105.045144,BandGeometry_Ahmed}.
\end{itemize} 

One should note that, for a given lattice model, in general the smallest bandwidth, nonzero Chern number, and the most ideal band geometry cannot be achieved simultaneously by optimizing the model parameters. In fact, it has been proven that one cannot get topological bands with zero bandwidth by finite-range hopping~\cite{Chen_2014}. Even if one introduces infinite-range hopping to get an exactly flat topological band~\cite{KapitMueller}, its Berry curvature still fluctuates~\cite{ZhaoKapit}. While perfectly flat band with exactly constant Berry curvature and finite Chern number can exist in multiband models (three or more bands) with infinite-range hopping, the trace and determinant conditions in Eq.~(\ref{fsb}) are not satisfied meanwhile~\cite{EmilBandGeometry}. For two-band models, both inequalities in Eq.~(\ref{fsb}) are saturated~\cite{BandGeometry_Lee,JieBandGeometry}, but the possibility of realizing exactly constant Berry curvature has been ruled out~\cite{EmilBandGeometry}. Therefore, we need to reach a compromise between the band dispersion, topology, and geometry when seeking ``good'' models to host FCIs. Over-optimizing one of those band properties does not always improve the stability of FCIs.

\subsection{Requirements for the interaction}
Generally speaking, interactions which decay fast with distance, like the onsite interaction (for bosons), Hubbard-like interactions between the nearest neighboring (NN) or next NN sites, long-range Coulomb and dipolar interactions, often work well to stabilize Abelian FCIs. However, for non-Abelian FCIs, one need to tune the interaction carefully, or even use less realistic multibody interactions. More rigorously, in analogy to what is done in the FQH case, we can use the lattice version of Haldane's pseudopotentials~\cite{HaldanePm} to predict what kind of interactions may stabilize a specific FCI state. These pseudopotentials on the lattice can be evaluated through the two-particle energy spectrum of the interaction~\cite{Andreas2013,zhao_review,ZhaoKapit} or the mapping between the LLL and the lattice Wannier functions~\cite{PPmChingHua}.

\subsection{Particle-hole asymmetry}
Having stated the requirements for the band and the interaction separately, we now turn to another key condition for the emergence of FCI states, whose satisfaction is determined jointly by the band and interaction properties. This condition originates from different behavior of the projected Hamiltonian Eq.~(\ref{projH}) under particle-hole (PH) transformation in a Bloch band from that in a LL. Under the standard PH transformation of fermions $\gamma_{\bf k}^\dagger \leftrightarrow\gamma_{\bf k}$, Eq.~(\ref{projH}) transforms to 
\begin{eqnarray}
H\rightarrow\sum_{{\bb k}\in {\rm 1BZ}} \epsilon_h({\bf k})\gamma^\dagger_{{\bb k}}\gamma_{{\bb k}}
+\sum_{\{ {\bf k}_i\}\in{\rm 1BZ}}\bar{V}^*_{\bb k_1\bb k_2\bb k_3\bb k_4}
\gamma^\dagger_{{\bb k_1}}\gamma^\dagger_{{\bb k_2}}\gamma_{{\bb k_3}}\gamma_{{\bb k_4}}
\label{projH2}
\end{eqnarray}
up to a constant, which includes a hole-hole interaction term similar to that of electrons and a single-hole dispersion
\begin{eqnarray}
\epsilon_h({\bf k})=-\epsilon({\bf k})
+\sum_{{\bf k}'\in{\rm 1BZ}}\left(\bar{V}_{\bb k \bb k' \bb k \bb k'}+\bar{V}_{\bb k' \bb k \bb k' \bb k}-\bar{V}_{\bb k \bb k' \bb k' \bb k}-\bar{V}_{\bb k' \bb k \bb k \bb k'}\right)
\equiv -\epsilon({\bf k})+\epsilon_h^i({\bf k}).
\label{Ehole}
\end{eqnarray}
Remarkably, apart from the contribution from the band dispersion $\epsilon({\bf k})$ of electrons, holes experience an additional dispersion $\epsilon_h^i({\bf k})$ originating from the interaction. In a LL, one can prove that $\epsilon_h^i$ is simply a constant shift (chemical potential) for two-body interactions~\cite{PhysRevB.72.045344}, which, together with the exact degeneracy of LL orbitals, means both electrons and holes in a LL experience the same zero dispersion (although this does not hold for multi-body interactions). However, $\epsilon_h^i({\bf k})$ generically varies with ${\bf k}$ in a Bloch band even for two-body interactions~\cite{Andreas2013,ZhaoKapit}, because the interaction matrix element $\bar{V}_{\bb k_1\bb k_2\bb k_3\bb k_4}$ in general lacks translational invariance in the reciprocal space of the lattice. This PH asymmetry, i.e., electrons and holes experience different dispersions, is a lattice-specific feature of the projected Hamiltonian in a Bloch band. As the strength of the PH asymmetry is quantified by the bandwidth of $\epsilon_h^i({\bf k})$, we need a weakly dispersive $\epsilon_h^i({\bf k})$ to simulate the LL physics on the many-body level. In fact, the bandwidth of $\epsilon_h^i({\bf k})$ is also closely related to the homogeneity of FS metric via
\begin{eqnarray} 
\epsilon_h^i({\bf k})\approx \sum_{\bf q}\tilde{V}({\bf q})e^{-\sum_{ab}q_a q_b g^{ab}({\bf k})}
\label{ehkfs}
\end{eqnarray} 
when the product of the interaction $\tilde{V}({\bf q})$ and the band form factor $\langle\psi({\bf k})|\psi({\bf k}-{\bf q})\rangle$ decay fast with $|{\bf q}|$~\cite{BandGeometry_Ahmed}. 

We can see the importance of uniform band geometry and weakly dispersive $\epsilon_h^i({\bf k})$ from Eq.~(\ref{ehkfs}). Similar to FQH states in a LL, FCIs prefers nearly uniform particle density and Berry curvature in the 1BZ, so that all particles can feel an approximately constant effective magnetic field. In narrow bands, the distribution of particles in the 1BZ is greatly affected by $\epsilon_h^i({\bf k})$: ${\bf k}$ points with small (large) $\epsilon^i_h({\bf k})$ tend to be occupied by holes (electrons). Therefore, if the FS metric is sufficiently uniform in a narrow band, the particles have no priorities to occupy the 1BZ, which, together with an approximately constant Berry curvature, should stabilize FCIs. Otherwise, strongly fluctuating FS metric will distribute electrons to regions with larger $\epsilon^i_h({\bf k})$ even when the bandwidth of $\epsilon({\bf k})$ is small. Such non-uniform density can destabilize the FCI phase, and can even completely destroy it if the Berry curvature happens to be weak in these regions~\cite{CDW_Patrick,BandGeometry_Ahmed}. We will discuss phases competing against FCIs caused by this mechanism in Sec.~\ref{sec:compete}. The crucial dependence of the many-body ground state on the band geometry is absent in LLs due to the uniform geometry there. In fact, flattening band geometry is one of the main challenges for realizing FCIs in experiments.

\section{Numerical diagnosis of FCIs}
\label{FCInumerics}
The conditions listed in the previous section serve as guiding principles to search for suitable candidate lattice models harboring FCIs. However, to quantitatively judge whether a candidate model is really good, we need to extract the low-energy physics of Eq.~(\ref{projH}) by numerical techniques such as exact diagonalization (ED) and density matrix renormalization group (DMRG). In this section, we will discuss how to diagnose FCIs by numerical simulations. We will give numerical identifications not only for FCIs, but also for some competing phases. A concrete example in twisted bilayer graphene will be presented in the next section.

\subsection{Basic identifications}

The standard numerical approach to explore the low-energy physics of Eq.~(\ref{projH}) is to diagonalize the many-body Hamiltonian for a partially filled band of a finite system, consisting of $N$ particles and $N_1$ and $N_2$ unit cells in the two primary directions of the lattice, then extract information from the obtained low-lying energy levels and eigenstates. Here the band filling is defined as $\nu\equiv N/(N_1 N_2)$, and interacting particles can either be electrons in solid materials or bosons in cold-atom setups. Some common methods for identifying FCIs' topological orders from numerical data are listed below. Depending on the method, the finite system is often put on the torus or the cylinder geometry. 

\begin{itemize}
\item 
{\bf Topological degeneracy.} The most obvious numerical evidence of FCIs is the topological degeneracy of incompressible (i.e., gapped) ground states on the torus, i.e., in a 2D system with periodic boundary conditions in both directions. This can be directly examined in the low-lying energy spectrum returned by numerical diagonalization, thus is usually the first thing to confirm in most literatures. In analogy to the conventional FQHE, at least $q$ approximately degenerate ground states are expected for FCIs at band filling $\nu=p/q$, where $p$ and $q$ are coprime (multicomponent and non-Abelian FCIs can have ground-state degeneracies larger than $q$). These ground states should carry momenta that match the Haldane statistics characterizing the candidate FCIs~\cite{FCIPRX,FCIsymmetry,HaldaneStatistics}. In general, the FCI ground-state degeneracies are not exact for finite systems due to the symmetry reduction compared to the conventional FQHE. However, the ground-state manifold should be separated from excited levels by a many-body gap that does not vanish in the thermodynamic limit and, ideally, the ground state splitting should vanish exponentially with the increasing of system size. Moreover, The ground-state degeneracy should be insensitive to the boundary conditions, which can be demonstrated by calculating the spectral flow under insertion of magnetic flux through the handle of the torus. Typically, the FCI ground states evolve into each other in the spectral flow for an appropriate flux insertion but do not mix with higher excited levels. 

\item 
 {\bf Entanglement measures.}  Besides energetic evidence, one can rely on entanglement spectroscopies for more insights into the nature of the ground states. Such entanglement information is entirely extracted from the numerically obtained ground-state wavefunctions. For a bipartition of a 2D topologically ordered state in real space, the entanglement entropy between the two subsystems obeys an area law, with a subleading term encoding the quantum dimensions of quasiparticles~\cite{TEEKitaev,TEELevin}. Numerical evaluation of this subleading term, dubbed topological entropy, for the ground state can distinguish FCIs from trivial phases~\cite{TEEFCI_Frank,PhysRevB.96.195123,TEEFCI_Titus,PhysRevB.101.235312}. Moreover, the entanglement spectrum (ES)~\cite{LiHaldane} of the ground state, i.e., the whole spectrum of the reduced density matrix of one subsystem, contains more information than the entanglement entropy. For the real-space bipartition, the counting structure in the low-lying ES~\cite{FCIRES_Zhao,TEEFCI_Titus,FCIPumping_Zhu,FCIRES_Frank,SS_Luo,FCIHighC_Luo,FCIRES_Gunnar,PhysRevB.101.235312} is the same as that of edge excitations~\cite{FCIEdge,FCIEdge2,PhysRevB.85.235137}, thus providing a fingerprint of the underlying topological order according to the bulk-edge correspondence. One can also bipartite the system by dividing all particles into two parts. The corresponding particle-cut ES (PES)~\cite{PES} encodes the physics of bulk quasihole excitations in its low-lying part, which has been extensively used to identify FCIs~\cite{FCIPRX,FCIWannier_Gunnar,FCIPES_Gunnar,FCIsymmetry,FCIZoology,FCIJain_Liu,FCIHighC_Zhao,FCIHighC_Sterdyniak,FCIorgan_Li,FCIHighC_Dong,FCIZ2_Regnault,ModelFCI_Zhao,FCIPES_Simon,FCICA_Sterdyniak,FCITBG_Ahmed,FCITDBG_Zhao,WC_Michal,FCITDTG_Wang}.

\item 
{\bf Hall conductance.}  As another hallmark, the quantized fractional Hall conductance is achieved when the ground state is in the FCI phase. The Hall conductance is proportional to the ground-state Chern number~\cite{Thouless,QianNiu1985}, which can be evaluated by twisting the boundary conditions on the torus~\cite{FCIChern_Wang,FCIChern_Wang2,PhysRevB.96.195123,FCIHalperin_Zeng,FCIHalperin_Zeng2,FCIHalperin_Zeng3,FCIHalperin_Zeng4,FCIHalperin_Zeng5,Okamoto2022}. Alternatively, for systems on the cylinder surface one can perform a numerical flux insertion through the hole of the cylinder and determine the Hall conductance by measuring the charges pumped from one edge to the other edge of the cylinder, which can be easily implemented in DMRG simulations~\cite{FCIPumping_Zaletel,TEEFCI_Frank,TEEFCI_Titus,FCIPumping_Zhu,FCIRES_Frank,FCIRES_Gunnar,FCIHole_Eckardt2,FCIHalperin_Zeng,FCIHalperin_Zeng2,FCIHalperin_Zeng3,FCIHalperin_Zeng4,FCIHighC_Luo,FCIHalperin_Zeng5,PhysRevB.101.235312}.

\item 
{\bf Quasiparticle statistics.} The fractional statistics of quasiparticles is a characterizing feature of FCI states. One can directly simulate the generation and braiding of quasiparticles to demonstrate their fractional statistics~\cite{FCIHole_Zhao,FCIHole_Anne,FCIHole_Eckardt,FCIHole_Anne2,FCIHole_Zhao2,FCIHole_Macaluso,FCIHole_Anne3,FCIHole_Eckardt2}. Alternatively, the statistics is encoded in the so called modular matrix~\cite{VERLINDE1988360,Wen1990}. We can numerically evaluate these matrices from the ground-state wavefunctions by constructing the minimally entangled states in the ground-state manifold for bipartitions of the lattice along different directions~\cite{MES_Zhang,FCIMES_Zhu,FCIMES_Zhu2,FCIPumping_Zhu,Gappedboundary_Zhao}.

\end{itemize}

\subsection{Competing phases}
\label{sec:compete}
FCI is only one possibility among various phases that can potentially exist in a Chern band. Competing phases prevail when the microscopic details of the system disfavor FCIs, which often happens when the conditions listed in Sec.~\ref{sec:conditions} are not satisfied. Studying the competition of FCIs with other candidate phases is crucial for determining the stable regions of FCIs in experiments. In the following, we discuss two common competing phases and their features which can be observed in numerical simulations.

\begin{itemize}
\item 
{\bf Crystalline phases.}
The competition in 2DEGs between conventional FQH states and charge density waves (CDWs), like Wigner crystals and stripe phases, has been known for long~\cite{WC_Anderson,WC_Lee,WC_Moessner,Stripe_Rezayi,CDW_Haldane,WC_Kun}. Both CDWs and FQH states rely on the presence of strong interactions, however, the formers spontaneously break the translational symmetry. Similarly, when the interaction dominates in topological flat bands, the competition between CDWs and FCIs is also typically expected at fillings that are commensurate with the lattice. Indeed, such competition has been confirmed in various flat-band models~\cite{FCIChern_Wang,CDW_Varney,CDW_Stefanos,CDW_Grushin,FCIorgan_Li,CDW_Stefanos2,CDW_Stefanos3,CDW_Stefanos4,SS_Luo,CDW_Patrick,WC_Jaworowski,WC_Michal,FieldTunedFCI,BandGeometry_Ahmed,FloquetFCI_Zhao}. Apparently, the formation of these states requires proper filling factor, interaction, and lattice shape. Recently, the studies of FCIs in moir\'e systems found that strongly fluctuating quantum geometry also facilitates the formation of CDWs by distributing particles to specific regions of the 1BZ~\cite{CDW_Patrick,FieldTunedFCI,BandGeometry_Ahmed}.

CDWs are characterized by their specific charge orders. In numerical simulations, this order can be revealed most prominently by calculating the structure factor  
\begin{eqnarray} 
S({\bf q})=\frac{1}{N_1 N_2}\left(\langle{\rho}({\bf q}) {\rho}(-{\bf q})\rangle-N^2\delta_{{\bf q},{\bf 0}}\right),
\label{sq}
\end{eqnarray} 
which is the Fourier transform of the static density-density correlation function. Here the average $\langle \rangle$ is over the ground state. When the charge order develops, $S({\bf q})$ should manifest pronounced peaks at the order momenta. These order momenta are reciprocal vectors of the real-space modulation of charge density. By contrast, $S({\bf q})$ of FCI states is expected to be featureless due to the absence of charge order. Regarding the low-energy many-body spectra, the CDW phase is accompanied by the ground-state degeneracy that corresponds to all possible charge configurations for a given charge order. Unlike the FCIs, the momenta of degenerate CDW states are separated by the order momenta instead of being determined by the Haldane statistics (occasionally these may coincide such that further diagnostics are needed to settle the nature of the states). 
 
\item
{\bf Fermi liquids.}
Generally speaking, there are two types of Fermi liquid (FL) that can emerge in a partially filled Bloch band. The first one is usual, caused by the strong dispersion of electron band $\epsilon({\bf k})$. This type of state can be identified by the Fermi surface structure in the dependence of electron's ground-state occupation $n({\bf k})$ on $\epsilon({\bf k})$ at momentum point ${\bf k}$, that is, $n({\bf k})$ suddenly drops from $\approx 1$ to $\approx 0$ at some critical $\epsilon({\bf k})$. The possibility of this usual FL phase can be ruled out when the electron band is very flat. 

The second type of FL phase is much less intuitive. As we showed in Eqs.~(\ref{projH2}) and (\ref{Ehole}), holes experience an additional dispersion $\epsilon_h^i({\bf k})$ induced by interactions, which may be strongly dispersive even in a narrow electron band. If $\epsilon_h^i({\bf k})$ dominates over the hole-hole interaction in a narrow electron band, FL becomes a good candidate for the ground state of holes, which still renders FL a natural ground-state candidate for electrons even if they do not feel strong band dispersion. Unlike the usual FL, this new FL is a correlated phase, as $\epsilon_h^i({\bf k})$ is contributed by the interaction in flat bands. It should be characterized by a strong correlation between $n({\bf k})$ and $\epsilon_h^i({\bf k})$: $n({\bf k})$ is small (large) at ${\bf k}$ points with small (large) $\epsilon_h^i({\bf k})$ which tend to be occupied (unoccupied) by holes. Also, the Fermi sea structure is expected to appear in the $n({\bf k})-\epsilon_h^i({\bf k})$ curve, which provides a fingerprint for the this unusual FL phase in numerical simulations. 

Intuitively, one may expect that the interaction-driven FL phase in narrow electron bands dominates over FCIs at high electronic band fillings. Indeed, this was confirmed in several toy models, where FCIs are replaced by the FL states at band filling $\nu\gtrsim 2/3$~\cite{Andreas2013,zhao_review}. However, in some moir\'e flat bands this unusual FL phase was found to prevail in a much wider range of $\nu$ down to $\nu=1/3$, where FCIs are totally absent~\cite{FCITBG_Ahmed,BandGeometry_Ahmed}. This can be understood by the very strong $\epsilon_h^i({\bf k})$ originating from the far-from-flat quantum geometry of these bands, as shown in Eq.~(\ref{ehkfs}).

\end{itemize}

\section{FCIs in twisted bilayer graphene}
\label{section:tbghbn}
After a general discussion of numerical identifications of FCI states and their competing phases, we now use twisted bilayer graphene (TBG) as a concrete example to show how these identifications work. Recent developments have picked moir\'e flat bands in TBG as promising platforms to host FCIs.

\subsection{Microscopic model of TBG}
TBG consists of two sheets of monolayer graphene with a twist angle $\theta$ in between~\cite{TBGReview_MacDonald}. At small twist angles $\theta\lesssim 10^\circ$, the difference between commensurate and incommensurate structures can be ignored, and the twisting leads to a moir\'e pattern of spatial periodicity $a_M=a/(2\sin(\theta/2))\gg a$ [Fig.~\ref{fig:TBG_Band}(a)], where $a=0.246 \ {\rm nm} $ is the lattice constant of monolayer graphene. The enlargement of the spatial periodicity corresponds to folding of original graphene 1BZ into much smaller moir\'e BZ (MBZ) [Fig.~\ref{fig:TBG_Band}(b)]. The low-energy states reside in the MBZs near the Dirac points ${\bf K}^{\pm}$ of monolayer graphene, called as two valleys of TBG which are related by time-reversal conjugation. At small twist angles, weak effects like intervalley scattering and spin-orbit coupling are negligible, leading to valley and spin degeneracy of each moir\'e band. The non-interacting low-energy band structure near the charge neutrality point (CNP) thus consists of four degenerate valence bands and four degenerate conduction bands, labeled by their valley index $+,-$ and spin index $\uparrow,\downarrow$. By convention the filling is chosen as $\nu=0$ at CNP, such that $\nu=-4$ ($\nu=+4$) when these eight bands are all empty (occupied). At a set of magic angles~\cite{Bistritzer2011,ChiralTBG_Ashvin}, valence and conduction bands become flat~\cite{Bistritzer2011,TBGContinuum_Santos,TBGContinuum_Santos2,ChiralTBG_Ashvin}, however, still touch each other. This Dirac band touching is protected by the combination of $C_{2z}$ rotation symmetry and time-reversal ($\mathcal{T}$) symmetry. On the single-particle level, the band gap can be opened by alignment with hexagonal boron nitride (hBN) which breaks the $\mathcal{C}_{2z} \mathcal{T}$ symmetry. In this case, the valence and conduction bands for each fixed valley and spin flavors are isolated and may carry opposite Chern numbers $\mathcal{C}=\pm 1$ in opposite valleys~\cite{TBGCBand_Hunt,TBGCBand_Jeil,PhysRevLett.123.036401,PhysRevB.99.155415,PhysRevB.99.195455,TBGCBand_Senthil,TBGCBand_Senthil2}, however, the valley and spin degeneracies of band energies remain. While it is complicated to predict how electrons occupy these degenerate bands when $\nu$ changes from $-4$ to $+4$~\cite{PhysRevLett.122.246401,Seo2019,PhysRevB.103.035427,PhysRevLett.124.187601,BernevigTBG-VI}, recent theoretical studies found that interactions are able to lift the valley and spin degeneracies under suitable circumstances~\cite{PhysRevLett.124.187601,PhysRevLett.124.166601,PhysRevB.103.035427,BernevigTBG-VI}. When this happens, we have a simple picture in which the eight spin and valley polarized bands are sequentially filled, giving rise to Chern insulating states at integer fillings. These Chern insulators at integer $\nu$ are indeed observed in experiments down to zero magnetic field~\cite{Zondiner2020,CI_Yazdani,CI_Wu,CI_Young,Das2021,CI_Choi,CI_Park,CI_Stepanov,Serlin2020,CI_Pierce}.

\begin{figure}
	\centering
	\includegraphics[width=\linewidth]{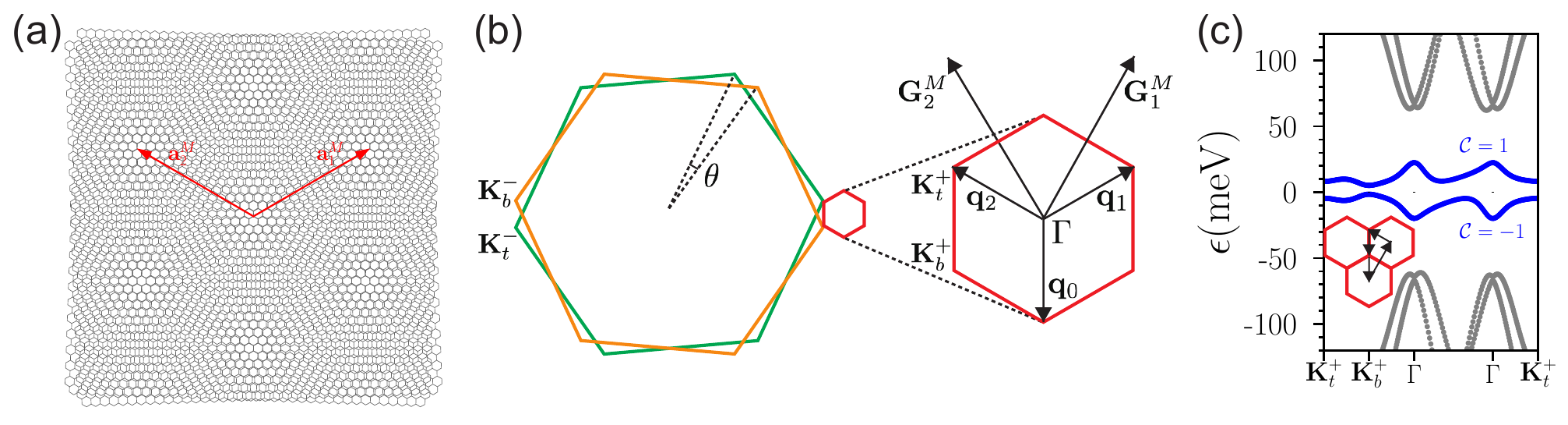}
	\caption{\label{fig:TBG_Band} (a) TBG moir\'e pattern in real space, whose periodicity is given by vectors ${\bf a}_1^M=a_M(\sqrt{3}/2,1/2)$ and ${\bf a}_2^M=a_M(-\sqrt{3}/2,1/2)$. (b) The moir\'e Brillouin zone (MBZ) of TBG. Two large hexagons (green and orange) are original first Brillouin zones of top and bottom graphene layers. The small red hexagon represents the MBZ resulting from twist. The ${\bf K}_{+,-}^{t,b}$ points, the vectors ${\bf q}_{0,1,2}$ and the MBZ reciprocal lattice vectors ${\bf G}_1$ and ${\bf G}_2$ are given. (c) Low-energy band structure of TBG-hBN for $\theta=1.30^\circ,M=20 \ {\rm meV}$ along a trajectory in the reciprocal space shown as the inset. The valence and conduction band (blue) near CNP carries Chern number $\mathcal{C}=-1$ and $\mathcal{C}=1$, respectively.}
\end{figure}

The above results motivate the investigation of interactions within a single band of TBG with fixed spin and valley flavors~\cite{FCITBG_Ahmed,TBGFCI_Ceceil,CDW_Patrick,FieldTunedFCI}. To proceed, we assume that the top and bottom graphene layers are rotated by $\pm\theta/2$, respectively, and hBN is aligned with the top layer. We focus on the valence band of valley ${\bf K}^+$. At small twist angles ($a_M\gg a$), the band information is encoded in an effective single-particle continuum Hamiltonian~\cite{Bistritzer2011,TBGCBand_Senthil,TBGCBand_Senthil2}
\begin{eqnarray}\label{eq:staticTBG}
			H_0 &=&  \sum_{\mathbf{k}} {\bf c}^{\dagger}_t(\mathbf{k}) h_{-\theta/2}(\mathbf{k}-{\bf K}^+_t) {\bf c}_{t}(\mathbf{k}) + \sum_{\mathbf{k}} {\bf c}^{\dagger}_b(\mathbf{k}) h_{\theta/2}(\mathbf{k}-{\bf K}^+_b) {\bf c}_{b}(\mathbf{k}) \nonumber\\
&+& \sum_{\mathbf{k}}\sum_{ j = 0}^2 \big({\bf c}_{t}^{\dagger}(\mathbf{k}-{\bf q}_0+\mathbf{q}_j) T_j {\bf c}_{b}(\mathbf{k}) + h.c. \big)  + M \sum_{\mathbf{k}} {\bf c}_{t}^{\dagger}(\mathbf{k}) \sigma_z {\bf c}_{t}(\mathbf{k}),
\end{eqnarray}
where ${\bf c}_{t/b}({\bf k}) = \begin{pmatrix}
c_{A_{t/b}}(\mathbf{k})\\ c_{B_{t/b}}(\mathbf{k})
\end{pmatrix}$ are spinors of annihilation operators for electrons with momentum ${\bf k}$ in top ($t$) and bottom ($b$) graphene layers, respectively, and $A$ and $B$ label the two sublattices of single-layer graphene. Note that ${\bf k}$ in $H_0$ is near the ${\bf K}^+$ valley, but not restricted in the MBZ. The first two terms in $H_0$ are just the standard Dirac Hamiltonians of top and bottom graphene layers, for which $h_{\theta}(\mathbf{k}) = h(R_{\theta} \mathbf{k}) $ with $h(\mathbf{k}) = -(\sqrt{3}a t_0/2)  (k_x \sigma_x + k_y \sigma_y)$, $t_0= 2.62\ {\rm eV}$, and $R_{\theta} $ the counter-clockwise rotation around the $z$-axis in the momentum space. To the lowest order of approximation, the alignment of hBN induces an onsite potential of strength $M$ (the last term in $H_0$) on the top graphene layer, where we have neglected the weaker effect from the moir\'e pattern formed between the top layer graphene and hBN~\cite{TBGCBand_Senthil2}. A rough estimation gives $M\approx 15 \ {\rm meV}$~\cite{TBGCBand_Senthil2}, however, larger values like $M=30 \ {\rm meV}$ were also used in theoretical studies~\cite{FCIexp_Xie,FieldTunedFCI}, so we choose $M=20 \ {\rm meV}$ throughout this section. The moir\'e tunneling between two graphene layers (the third term in $H_0$) is given by
$T_j = w_0 - w_1 e^{i(2\pi/3)j \sigma_z} \sigma_x e^{-i(2\pi/3)j\sigma_z}$
with $\mathbf{q}_0 =  \mathbf{K}^+_t - \mathbf{K}^+_b$, $\mathbf{q}_1  = R_{2\pi/3} \mathbf{q}_0$ and $\mathbf{q}_2 = R_{-2\pi/3} \mathbf{q}_0$ [Fig.~\ref{fig:TBG_Band}(b)], where $w_0$ and $w_1$ are the interlayer tunneling strengths between $AA$ and $AB$ sites, respectively. $w_0$ is smaller than $w_1$ because of lattice relaxation~\cite{Popov2011,Nam2017,Uchida14,Wijk_2015,Koshino2018}. Theoretical calculations and experimental measurements suggest $w_0/w_1=0.5-0.8$~\cite{Koshino2018,Xie2019,Francisco2019,Das2021}, hence we choose $w_0=0.7w_1$ in this section, where $w_1$ is fixed as $110 \ {\rm meV}$ according to {\em ab initio} numerics~\cite{Jung14}.  For each ${\bf k}_0\in{\rm MBZ}$, the moir\'e band structure in the MBZ can be obtained by writing ${\bf k}$ in $H_0$ as ${\bf k}={\bf k}_0+m{\bf G}_1+n{\bf G}_2$ with integers $m,n=-d,...,d$ then diagonalizing $H_0$, where ${\bf G}_1$ and ${\bf G}_2$ are the MBZ primitive reciprocal lattice vectors [Fig.~\ref{fig:TBG_Band}(b)] and $d$ is a cutoff typically set as $d=5-9$. 
In Fig.~\ref{fig:TBG_Band}(c), we show the low-energy band structure at $\theta=1.30^\circ$, where the flat valence and conduction bands near CNP have Chern number $\mathcal{C}=\mp 1$, respectively.

Regarding the interaction, we consider a dual-gate setup to mimic experiments, in which electrons experience the gate-screened Coulomb interaction, whose Fourier transform reads
\begin{eqnarray}\label{eq:screenedcoulomb}
\tilde{V}({\bf q})=\frac{e^2}{4\pi\varepsilon A}\frac{2\pi\tanh(\xi q/2)}{q},
\end{eqnarray}
where $\varepsilon$ is the dielectric constant of the material and $\xi$ is the distance between the top and bottom gates. Based on typical parameters in TBG experiments, we choose $\varepsilon=4\varepsilon_0$~\cite{C8NA00350E,Guinea18} with $\varepsilon_0$ the vacuum dielectric constant and scan $\xi$ between $a_M$ and $2a_M$~\cite{Saito2020,Stepanov2020}. 

The total Hamiltonian projected to the valence band has the same form as Eq.~(\ref{projH}), in which ${\bf k}$'s should be now restricted in the MBZ and $\epsilon_{\bf k}$ is the valence band dispersion. The matrix elements of the projected interaction are 
\begin{eqnarray}\label{eq:TBGV}
\bar{V}_{\bb k_1\bb k_2\bb k_3\bb k_4}=\frac{1}{2}\delta'_{\bb k_1+\bb k_2,\bb k_3+\bb k_4}\sum_{\bf G} \tilde{V}(\bb k_1-\bb k_4+\bb G)\langle \psi({\bf k}_1)|\psi({\bf k}_4-\bb G)\rangle
\langle \psi({\bf k}_2)|\psi({\bf k}_3+\bb G+\delta\bb G)\rangle,
\end{eqnarray}
where ${\bf G}$ is summed over the entire reciprocal space instead of MBZ only, $\delta\bb G=\bb k_1+\bb k_2-\bb k_3-\bb k_4$, and $|\psi({\bf k})\rangle$ is the valence band eigenvector. One may further account the effects of remote bands in the mean-field level by adding extra single-particle terms~\cite{BernevigTBG-III,BernevigTBG-VI,FieldTunedFCI}, however, we neglect this issue to avoid technical complexity in this tutorial introduction.  

\begin{figure}
	\centering
	\includegraphics[width=\linewidth]{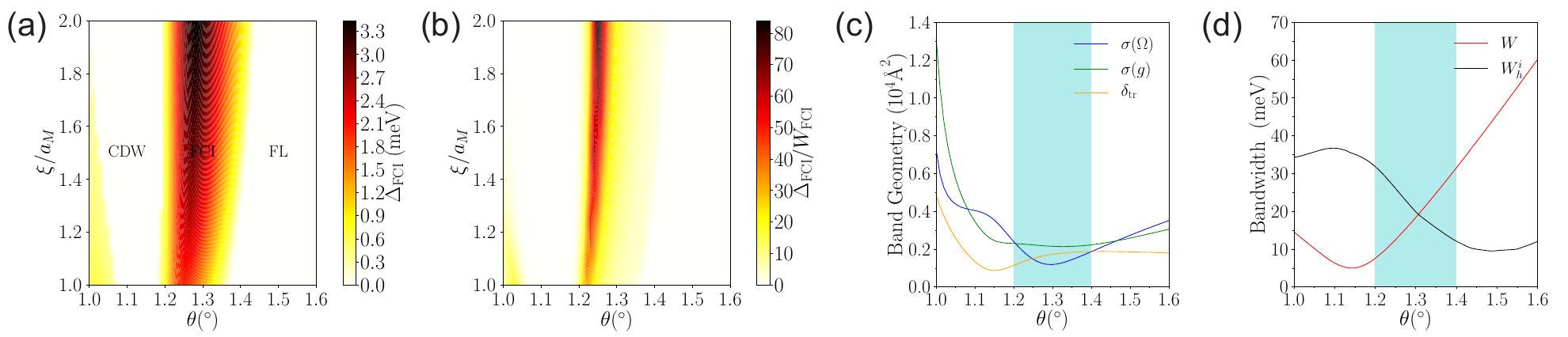}
	\caption{\label{fig:TBG_PD} Numerical data in the $\theta-\xi$ parameter space of TBG-hBN. (a) The FCI gap $\Delta_{\rm FCI}$ and (b) the quality of FCI degeneracy $\Delta_{\rm FCI}/W_{\rm FCI}$ for $N=10$ electrons in the valence band of $N_1\times N_2=5\times 6$ moir\'e lattice. (c) Band geometry of the valence band, including the Berry curvature fluctuation $\sigma(\Omega)$, the FS metric fluctuation $\sigma(g)$, and the deviation $\delta_{\rm tr}$ from the saturated trace condition. (d) Bandwidths of the valence band ($W$) and the interaction-induced hole dispersion $(W_h^i)$. The shades in (c) and (d) indicates the range of twist angle in which robust FCIs exist. In (a), we give a tentative phase diagram. As shown later, the CDW phase is identified by the structure factor, and the usual Fermi liquid (FL)-like phase is characterized by the step structure in the $n({\bf k})-\epsilon({\bf k})$ curve.  }
\end{figure}

\subsection{Numerical evidence of FCIs}

With the microscopic many-body Hamiltonian projected to the valence band of the ${\bf K}^+$ valley, we can run extensive ED calculations on the torus geometry to search for numerical identifications of FCI states, as discussed in Sec.~\ref{FCInumerics}. We do this in the parameter space spanned by $\theta\in[1.0^\circ,1.6^\circ]$ and $\xi/a_M\in[1,2]$, in which the valence band is always isolated and carries Chern number $\mathcal{C}=-1$. The system has a finite size, consisting of $N$ electrons on the $N_1\times N_2$ moir\'e lattice. Considering that robust FCIs at one-third band filling were observed in various $|\mathcal{C}|=1$ flat bands~\cite{Sheng2011,FCIPRX,FCIZoology,Andreas2013}, we choose $\bar\nu\equiv N/(N_1N_2)=1/3$ filling of the valence band and expect the emergence of lattice analogs of the Laughlin FQH state~\cite{Laughlin83}. 

First of all, provided all eigenvalues $\{ E_n\}$ of the many-body Hamiltonian returned by ED have been sorted in ascending order, the three-fold degeneracy of incompressible FCI ground states can be probed by the FCI gap $\Delta_{\rm FCI}\equiv E_4-E_1$ and the FCI splitting $W_{\rm FCI}\equiv E_3-E_1$. If the lowest three eigenstates are not in momentum sectors predicted by the Haldane statistics of the Laughlin state, we simply set $\Delta_{\rm FCI}=0$. The numerical data of scanning $\Delta_{\rm FCI}$ and $\Delta_{\rm FCI}/W_{\rm FCI}$ in the $(\theta,\xi)$ parameter space for $N = 10$ electrons are demonstrated in Figs.~\ref{fig:TBG_PD}(a) and \ref{fig:TBG_PD}(b), from which we can identify a wide range of parameters supporting nice three-fold degeneracies in the momentum sectors of the Laughlin state. This region slightly expands in the $\theta$ direction from $\theta\in[1.20^\circ,1.35^\circ]$ to $\theta\in[1.20^\circ,1.40^\circ]$ with the increasing of $\xi$ from $a_M$ to $2a_M$ (weaker screening). 

Having obtained the three-fold degeneracy for $N = 10$ electrons as preliminary evidence of the $\bar\nu=1/3$ FCI, we analyze the numerical data more carefully at a representative parameter point $(\theta,\xi)=(1.27^\circ,2a_M)$ in this region (Fig.~\ref{fig:TBG_FCI}). First, the dependence of the degeneracy on the system size and the boundary condition should be examined. The energy spectra of different system sizes, plotted versus the many-body momentum $(K_1,K_2)$, are presented in Fig.~\ref{fig:TBG_FCI}(a), in which the three-fold degeneracy in the momentum sectors predicted for the Laughlin state appear for all system sizes. Moreover, this degeneracy persists under magnetic flux insertion through the handles of the toroidal system [Fig.~\ref{fig:TBG_FCI}(c)]. A finite-size scaling of $\Delta_{\rm FCI}$ and $W_{\rm FCI}$ suggests that both the three-fold degeneracy and the ground-state gap are very likely to survive in the thermodynamic limit [Fig.~\ref{fig:TBG_FCI}(b)]. Note that the gap extrapolating to the $1/N\rightarrow 0$ limit goes to $\sim2 \ {\rm meV}$, corresponding to a temperature which is an order of magnitude higher than required by the conventional FQHE in 2DEGs. The nontrivial topological properties of the ground state can be further corroborated by the PES, whose levels $\{\lambda_n\}$ are labeled by the total momentum $(K_1^A, K_2^A)$ of the subsystem with $N_A$ electrons. Strikingly, an entanglement gap appears in the PES [Fig.~\ref{fig:TBG_FCI}(d)], the number of levels below which exactly matches the counting of quasihole excitations in the Laughlin state~\cite{FCIPRX}.

\begin{figure}
	\centering
	\includegraphics[width=\linewidth]{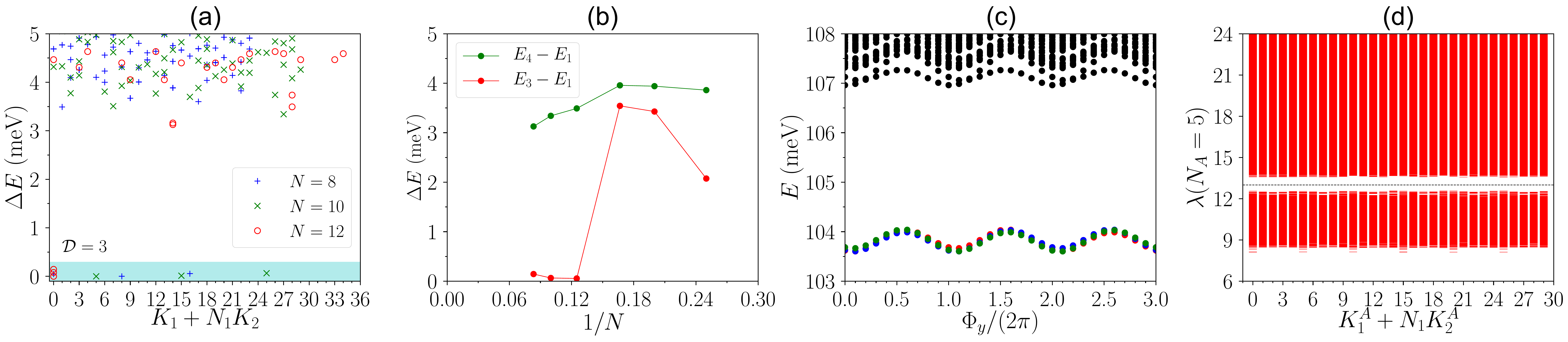}
	\caption{\label{fig:TBG_FCI} Numerical evidence of the $\bar\nu=1/3$ FCI in the valence band of TBG-hBN at a representative parameter point $(\theta,\xi)=(1.27^\circ,2a_M)$. (a) The low-lying energy spectra for $N=8,10,12$ electrons on the $N/2\times 6$ lattice. The shade indicates the three-fold ($\mathcal{D}=3$) topological degeneracy of the ground states. (b) The finite-size scaling of the energy gap $E_4-E_1$ and the ground-state splitting $E_3-E_1$ for $N=4, 5, 6, 8, 10,12$ electrons. (c) The spectral flow for $N=10$, $N_1\times N_2=5\times 6$, where $\Phi_y$ is the magnetic flux insertion in the ${\bf a}_2^M$ direction. (d) The particle entanglement spectrum for $N=10$, $N_1\times N_2=5\times 6$ and $N_A=5$, with $23256$ levels below the entanglement gap (the dashed line).}
\end{figure}

Both the low-energy spectrum and entanglement spectroscopy strongly indicate that lattice analogs of the continuum $\nu=1/3$ Laughlin state exist in the valence band of TBG-hBN. We can even establish their adiabatic continuity by using the mapping between single-particle states in a Chern band and in the LLL. Such mapping has been constructed either between lattice Wannier states and LLL orbitals in the Landau gauge~\cite{Wannier_Qi,Wannier_Wu,Wannier_Qi2,HaldaneStatistics}, or between lattice Bloch states and LLL Bloch-like basis~\cite{FCIHighC_YangLe}. In fact, TBG in the so called chiral limit~\cite{ChiralTBG_Ashvin} allows to build a more explicit correspondence between $|\mathcal{C}|=1$ Chern bands with {\em ideal geometry} and the LLL. To reach this limit, we need to set $w_0=0$ such that the single-particle Hamiltonian possesses chiral symmetry $\sigma_z H_0 \sigma_z=-H_0$, where $\sigma_z$ is the Pauli matrix acting on the sublattice. While it is a drastic approximation to decrease $w_0$ to zero even in the presence of lattice relaxation, the magic angle phenomena of TBG can already be well captured in the chiral limit: the entire valence and conduction bands near CNP become perfectly flat at a set of twist angles when $w_0=0$, which is the origin of the nearly flat bands observed at magic angles with finite $w_0$~\cite{ChiralTBG_Ashvin}. In the presence of chiral symmetry, the magic-angle TBG (MATBG) flat bands gains ideal quantum geometry~\cite{ChiralTBG_Ashvin2}, that is, the Berry curvature does not change sign and varies in sync with the FS metric via 
\begin{eqnarray} 
g^{ab}({\bf k})=\frac{1}{2}\delta^{ab}|\Omega({\bf k})|,
\label{fsbtbg}
\end{eqnarray} 
in the entire MBZ, where $\delta^{ab}$ is the identity matrix. In this situation, both inequalities in Eq.~(\ref{fsb}) are saturated, just like in the LLL (see Ref.~\cite{PhysRevB.104.115160} for a systematic method to construct bands with ideal quantum geometry). Moreover, the real-space forms $\psi_{\bf k}({\bf r})$ of flat-band wave functions in chiral MATBG are fixed by the ideal quantum geometry as~\cite{JieBandGeometry}
\begin{eqnarray} 
\psi_{\bf k}({\bf r})=\mathcal{N}_{\bf k}\mathcal{B}({\bf r})\varphi_{\bf k}({\bf r}),
\label{tbgpsi}
\end{eqnarray} 
where $\mathcal{N}_{\bf k}$ is the normalization factor, $\mathcal{B}({\bf r})$ is a ${\bf k}$-independent function, and $\varphi_{\bf k}({\bf r})$ is the Bloch-like basis in the LLL. Remarkably, Eq.~(\ref{tbgpsi}) does not only hold for chiral MATBG, but is valid for all $|\mathcal{C}|=1$ bands with the ideal quantum geometry~\cite{JieBandGeometry}. Starting from this relation in the single-particle level, one can also generalize Haldane pseudopotentials from the LLL to ideal $|\mathcal{C}|=1$ flat bands and construct model FCIs as zero-energy ground states of short-ranged interactions~\cite{ChiralTBG_Ashvin2,JieBandGeometry}. Zero-energy FCIs were also observed in other models~\cite{KapitMueller,ZhaoKapit,ModelFCI_Zhao}, which implies that the band geometry is also ideal there. The ideal band geometry and model FCIs are also expected in chiral MATBG at a finite magnetic field~\cite{ChiralTBG_Ady}.

Back to our numerical results in TBG-hBN, the stabilization of the $\bar\nu=1/3$ FCI phase is closely related to whether the conditions listed in Sec.~\ref{sec:conditions} are satisfied. To see this, we characterize quantum geometry of the valence band by fluctuations of the Berry curvature $\Omega({\bf k})$ and FS metric $g^{ab}({\bf k})$ and the violation of the saturated trace condition, as quantified by
\begin{eqnarray} 
\sigma(\Omega)&=&\sqrt{\langle\Omega^2({\bf k})\rangle-\langle\Omega({\bf k})\rangle^2}, \nonumber\\
\sigma(g)&=&\sqrt{\sum_{a,b=x,y}\left(\langle g^{ab}({\bf k}) g^{ba}({\bf k})\rangle-\langle g^{ab}({\bf k})\rangle \langle g^{ba}({\bf k})\rangle\right)},\nonumber\\
\delta_{\rm tr}&=&\langle{\rm tr}g^{ab}({\bf k})-|\Omega({\bf k})|\rangle,
\label{sigma}
\end{eqnarray} 
respectively, where $\langle O({\bf k})\rangle=\frac{1}{A_{\rm MBZ}}\int_{\rm MBZ} O({\bf k}) dk_x dk_y$ is the average of quantity $O$ in the MBZ, with $A_{\rm MBZ}$ the area of MBZ. We plot $\sigma(\Omega)$, $\sigma(g)$, $\delta_{\rm tr}$, together with the bandwidth $W$ of the electron's valence band and the bandwidth $W_h^i$ of interaction-induced hole dispersion $\epsilon_h^i({\bf k})$ in Figs.~\ref{fig:TBG_PD}(c) and \ref{fig:TBG_PD}(d) as a function of twist angle $\theta$. Remarkably, the most robust FCI states indeed appear when all of these quantities are sufficiently small. Nevertheless, the minima of these quantities, are not located at the same $\theta$, so over-minimizing one quantity does not necessarily improve the stability of the FCI phase. In particular, the robust FCIs appear at twist angles slightly above the TBG magic angle. These numerical results are consistent with the requirements listed in Sec.~\ref{sec:conditions}. 

\subsection{Numerical evidence of competing phases}
As we discussed before, competing phases will prevail when properties of the microscopic model disfavor FCI states. In our numerical data, the FCI phase obviously breaks down when the twist angle is either reduced or increased out of the optimal range in Fig.~\ref{fig:TBG_PD}. On the left side of the FCI phase (with smaller $\theta$), there is a pronounced sensitivity of the energy spectrum to the lattice size. In particular, when both $N_1$ and $N_2$ are divisible by three, we observe a new kind of three-fold ground-state degeneracy appearing in equally spaced momentum sectors rather than in $\bar\nu=1/3$ Laughlin FCI sectors. For example, at a representative parameter point $(\theta,\xi)=(1.10^\circ, 2a_M)$ in this region, the three ground states of $N=12,N_1\times N_2=6\times 6$ appear in the $(K_1,K_2)=(0,0), (2,2)$ and $(4,4)$ sectors [Fig.~\ref{fig:TBG_compete}(a)]. By contrast, the ground states in the FCI phase of the same system size are all located in the $(K_1,K_2)=(0,0)$ sector [Fig.~\ref{fig:TBG_FCI}(a)]. 

\begin{figure}
	\centering
	\includegraphics[width=\linewidth]{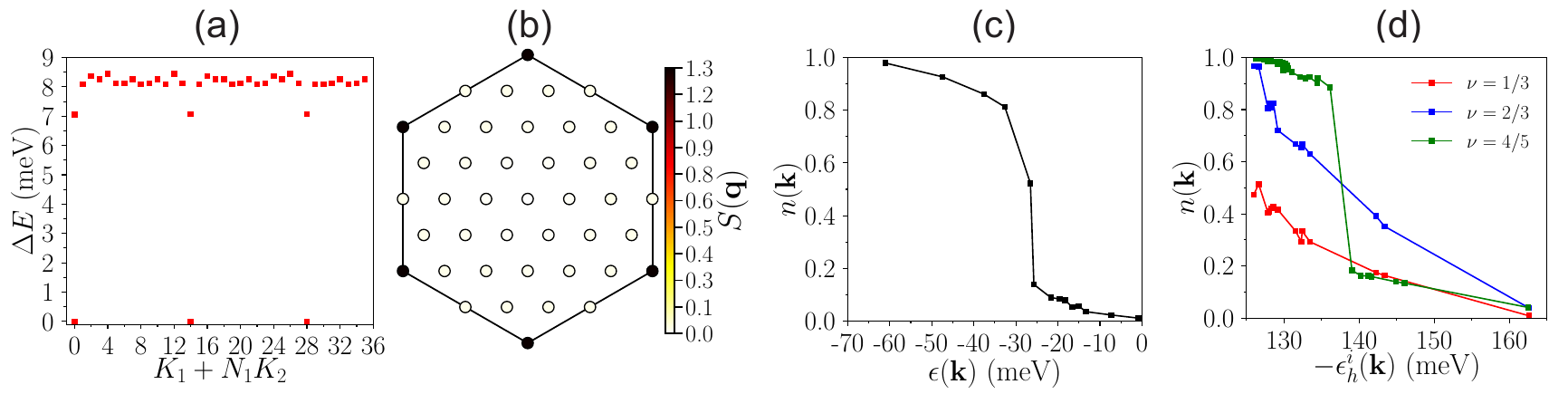}
	\caption{\label{fig:TBG_compete} Numerical evidence of competing phases in the valence band of TBG-hBN. (a) The low-lying energy spectrum of the CDW phase at a representative parameter point $(\theta,\xi)=(1.10^\circ,2a_M)$. The three ground states at separated by the order momenta $\Delta{\bf K}=(2,2)$ and $(4,4)$. (b) The structure factor $S({\bf q})$ of the CDW state at the same parameter point as in (a). (c) The ground-state occupation $n({\bf k})$ of electrons versus the electron band dispersion in the usual FL phase at a representative parameter point $(\theta,\xi)=(1.60^\circ,2a_M)$. We choose $N=12$ electrons at $\bar\nu=1/3$ in the $N_1\times N_2=6\times 6$ lattice in (a)-(c). (d) The ground-state occupation $n({\bf k})$ of electrons versus the interaction-induced hole dispersion $\epsilon_h^i({\bf k})$ at the parameter point $(\theta,\xi)=(1.10^\circ,2a_M)$ for fillings $\bar\nu=1/3,2/3$ and $4/5$. The corresponding system sizes are $N=12,N_1\times N_2=6\times 6$, $N=24,N_1\times N_2=6\times 6$, and $N=28,N_1\times N_2=5\times 7$, respectively.}
\end{figure}

Distribution of degenerate ground states over equally spaced momenta is a signal of CDWs. As we discussed in Sec.~\ref{sec:compete}, this can be further confirmed by the structure factor $S({\bf q})$. Indeed, pronounced peaks of $S({\bf q})$ exist at the corners of the MBZ [Fig.~\ref{fig:TBG_compete}(b)], revealing the underlying phase is the CDW with the order momentum ${\bf K}^+_{t,b}$. These order momenta are indeed those separating the degenerate ground states, by noting $\Delta{\bf K}=(2,2)=({\bf G}_1+{\bf G}_2)/3\sim{\bf K}^+_b$ and $\Delta{\bf K}=(4,4)=2({\bf G}_1+{\bf G}_2)/3\sim{\bf K}^+_t$. Here $\sim$ means ``equal to'' up to a MBZ primitive reciprocal lattice vector. The order momenta suggest a tripled unit cell of charge modulation compared to the original moir\'e lattice~\cite{CDW_Patrick}. The identification of order momenta also explains the absence of the CDW phase when either $N_1$ or $N_2$ are indivisible by three. Remember that the single-electron momentum ${\bf k}$ can only take ${\bf k}=\frac{m_1}{N_1}{\bf G}_1+\frac{m_2}{N_2}{\bf G}_2$ for a finite periodic system. Hence both $N_1$ and $N_2$ must be divisible by three if ${\bf K}^+_{t,b}$ belong to this set of allowed ${\bf k}$. 

On the other hand, on the right side of the FCI phase (larger $\theta$), neither the FCI degeneracy nor the CDW degeneracy appears in the low-energy spectra, no matter what the lattice size is. Note that in this region the electron bandwidth dramatically increases while the interaction-induced hole bandwidth remains small [Fig.~\ref{fig:TBG_PD}(d)], implying that the ground state is a usual FL. Indeed, we find a sharp step in the electronic occupation number $n({\bf k})$ at some critical $\epsilon({\bf k})$, as shown in Fig.~\ref{fig:TBG_compete}(c). 

To see the unusual interaction-induced FL phase, we need to check the region in which the interaction-induced hole dispersion is much stronger than electron dispersion. Therefore, we choose the parameter point $(\theta,\xi)=(1.10^\circ,2a_M)$ where this condition is satisfied [Fig.~\ref{fig:TBG_PD}(d)]. Remarkably, we do find the emergence of a Fermi-surface structure in the $n({\bf k})-\epsilon_h^i({\bf k})$ with the increasing of the filling factor from $1/3$ to $4/5$ [Fig.~\ref{fig:TBG_compete}(d)]. This means the ${\bf K}$-CDW phase at $\bar\nu=1/3$ is replaced by the unusual FL phase at large fillings.

\section{FCIs in Bloch bands with high Chern numbers}
As listed in Table~\ref{tbl1}, a significant qualitative difference of a Chern band from a LL is that it can support higher Chern number $|\mathcal{C}|>1$. In general, one can add longer-range hopping in a $|\mathcal{C}|=1$ model or appropriately stack multilayers of $|\mathcal{C}|=1$ models to get $|\mathcal{C}|>1$ bands~\cite{WangFa2011,PhysRevLett.106.156801,Trescher2012,Yang2012,Sticlet2013,Wu2015,ModelFCI_Zhao,TBGCBand_Senthil,TLGC3,AshvinTDBG}. This strategy also works for moir\'e materials. One can systematically construct $|\mathcal{C}|=n$ bands near the CNP by twisting two sheets of Bernal-stacked $n$ graphene layers~\cite{FCITDTG_Wang,FCITDTG_Ashvin}. TBG corresponds to the $n=1$ case. $n=2$ gives the twisted double bilayer graphene (TDBG), for which $|\mathcal{C}|=2$ flat bands near the CNP have been confirmed~\cite{TBGCBand_Senthil,AshvinTDBG,TDBG_Jung}. Strikingly, the model constructed in this way gains chiral symmetry for any $n\geq 1$ if only intersublattice couplings are kept. At magic angles, these chiral models all possess perfectly flat valence and conduction bands with ideal band geometry. 

FCIs can be diagnosed in high-Chern number bands by the same identifications as in $|\mathcal{C}|=1$ models. Excitingly, series of high-$\mathcal{C}$ FCI states have been numerically discovered in realistic moir\'e bands~\cite{FCITDBG_Zhao,FCITDTG_Wang} inspired by earlier numerical findings in tight-binding toy models~\cite{Yang2012,YiFei2012,FCIHighC_Zhao,FCIHighC_Dong,FCIHighC_Sterdyniak,FCIHighC_YangLe,FCIHighC_Emil,Wu2015,ModelFCI_Zhao,GunnarHof,PhysRevB.97.035159}. These FCIs appear at band fillings, for instance, 
\begin{eqnarray} 
\nu=\frac{k}{|\mathcal{C}|(m-1)+1}
\label{highc_fci}
\end{eqnarray} 
with integer $k\geq1$ and $m\geq 2$. Here $k=1$ and $k>1$ correspond to Abelian and non-Abelian states, respectively, and $m$ is even (odd) for bosons (fermions). Further fillings of high-$\mathcal{C}$ FCIs can be predicted by applying the composite-fermion theory to Hofstadter bands~\cite{GunnarHof}.
Some of the observed high-$\mathcal{C}$ FCIs are even model states occurring as the zero-energy ground states of proper short-ranged interactions~\cite{ModelFCI_Zhao,FCITDTG_Wang}.

Ample attention has been directed toward understanding these high-$\mathcal{C}$ FCIs. Intuitively, they may have a relation with the conventional FQH states in $|\mathcal{C}|$ copies of the LLL, which also carry Chern number $\mathcal{C}$ as a whole. However, an obvious discrepancy exists between a single Chern number $\mathcal{C}$ band and $|\mathcal{C}|$ copies of the LLL: the number of single-particle orbitals in the latter must be divisible by $|\mathcal{C}|$ as each LLL contains an integer number of orbitals, while the number of single-particle states in the former does not necessarily to be a multiple of $|\mathcal{C}|$. To resolve this discrepancy, a new type of $|\mathcal{C}|$-component LLL was proposed. Here component refers to internal degrees of freedom of particles, like spin or physical layer. In contrast to the usual multicomponent LLL where the $|\mathcal{C}|$ components are decoupled, a new set of boundary conditions, called ``color-entangled'' boundary conditions, is adopted to connect different components, i.e., a particle can change its component index when crossing the boundary of the system~\cite{FCIHighC_YangLe,HaldaneStatistics}. Under this scenario, one can construct a single manifold of Bloch-like states with Chern number $\mathcal{C}$, rather than $|\mathcal{C}|$ separate Chern-number-one manifolds, such that the number of orbitals per LLL does not need to be an integer. Diagonalizing Haldane's pseudopotentials in the ``color-entangled'' LLL basis produces unconventional FQH states, which were found to have high overlaps and the same PES countings with high-$\mathcal{C}$ FCIs in lattice models~\cite{FCIHighC_YangLe}. However, the analytical ansatz wavefunctions of the high-$\mathcal{C}$ FCIs are still unknown, and it remains unclear whether the color-entangled boundary conditions lead to new topological orders compared with conventional $|\mathcal{C}|$-component FQH states. 

\section{Experimental observation of FCIs}
We have seen that FCIs arise from the interplay between interactions and the dispersion, topology, and geometry of the partially filled band. Therefore, their experimental realizations likely require highly controllable platforms in which band properties and interactions can be easily tuned to the suitable regime. Given this, van-der-Waals heterostructures with moir\'e patterns~\cite{Geim2013,Novoselov2016} naturally emerge as a promising candidate to host FCIs. There are many experimental knobs in this kind of setups that can affect band properties and interactions, including the twist angle, the external fields, the dielectric environment, and even the material itself. In the past few years, rapid progress has been made on manufacturing van-der-Waals heterostructures and investigating the exotic correlated phases thereof.

\begin{table}
\caption{The observed incompressible states in Refs.~\cite{FCI_Eric,FCIexp_Xie} classified by their values of $t$ and $s$. While some of the observed Chern insulators and FCIs preserve translational symmetry (TS) of the moir\'e superlattice, others spontaneously break this symmetry with an expanded unit cell of electron density distribution, implying an intriguing combination of conventional charge density wave and topological features. Those exotic fractional Chern insulators in the last row are neither simple TS preserving nor TS broken states, and may originate from the interplay between the multicomponent nature of the material and the spatial symmetry.}
\label{tbl2}
\begin{tabular*}{\tblwidth}{m{6cm}m{10cm}m{10cm}c}
\toprule
Incompressible states \centering &  Values of $t$ and $s$ \centering &  \\ % Table header row
\midrule
trivial correlated insulator \centering & $t=0$ and integer $s\neq 0$ \centering & \\
\hline
charge density wave  \centering& $t=0$ and fractional $s$ \centering & \\
\hline
integer quantum Hall states in LLs  \centering& integer $t\neq 0$ and $s=0$ \centering &\\
\hline
TS-preserving Chern insulators  \centering& integer $t\neq 0$ and integer $s\neq 0$ \centering &\\
\hline
TS-broken Chern insulators (also observed in Refs.~\cite{CI_Pierce,Polshyn2022})  \centering& integer $t\neq 0$ and fractional $s$ \centering & \\
\hline
fractional quantum Hall states in LLs  \centering& fractional $t$ and $s=0$ \centering &\\
\hline
TS-preserving fractional Chern insulators  \centering& fractional $t$ and fractional $s$, $t$ and $s$ have the same denominator \centering & \\
\hline
TS-broken fractional Chern insulators \centering & fractional $t$ and fractional $s$, denominator of $s$ is a multiple of that of $t$ \centering & \\
\hline
exotic fractional Chern insulators  \centering& fractional $t$ and fractional $s$, $t$ and $s$ have coprime denominators \centering & \\
\bottomrule
\end{tabular*}
\end{table}

\subsection{Experiment in Bernal-stacked bilayer graphene}

The first observation of FCI states was reported in a van-der-Waals heterostructure made of Bernal-stacked bilayer graphene aligned with hBN (BLG-hBN), penetrated by a perpendicular magnetic field $B\sim 30 \ {\rm T}$~\cite{FCI_Eric}. In this system, electrons experience a moir\'e superlattice potential from the mismatched graphene and hBN lattices. The interplay between the magnetic field and the moir\'e potential creates topological Harper-Hofstadter bands that act as parents of potential FCI states. By determining the compressibility of the system via measuring the penetration field capacitance for variable electron density $n_{\rm e}$ and magnetic flux density $n_\phi$ per moir\'e superlattice cell, multiple incompressible states are identified along trajectories obeying Streda formula
\begin{eqnarray} 
n_{\rm e}=tn_\phi+s,
\label{streda}
\end{eqnarray} 
where $t=\sigma_{xy}h/e^2$ is the dimensionless Hall conductance and $s$ is the zero-field filling factor~\cite{Streda_1982,PhysRevB.28.6713}. 

The observed incompressible states can be distinguished by their values of $t$ and $s$ (see Table~\ref{tbl2}): non-integer $t$ signals intrinsic topological order whereas fractional $s$ may originate from either topological order or translational symmetry (TS) breaking. Remarkably, apart from the trivial states, QHE states in LLs, and Chern insulators in Hofstadter bands, clear evidence of states with fractional $s$ and fractional $t$ was observed in the experiment, which correspond to FCIs. These FCI states develop from electron/hole doping the nearby parent Chern insulating states with an effective filling $\bar\nu$ in the partially filled band. Some of them exist in a $\mathcal{C}=-1$ band at $\bar\nu=1/3,2/3,2/5,3/5$, falling into the odd-denominator Laughlin and composite fermion sequence. Notably, several other states exist in a $\mathcal{C}=2$ band, for instance, at $\bar\nu=1/3$ and $\bar\nu=1/6$. The $\mathcal{C}=2,\bar\nu=1/3$ state, with $(t,s)=(8/3,-1/3)$, preserves TS of the moir\'e superlattice and belongs to the sequence predicted in Ref.~\cite{GunnarHof}. However, the $\mathcal{C}=2,\bar\nu=1/6$ state carries $(t,s)=(7/3,-1/6)$, which means that only half fundamental charge $e/3$ is bound to each moir\'e unit cell. This feature suggests doubling of unit cell in the electron density distribution which spontaneously breaks TS of the moir\'e superlattice. The TS-broken FCI, as well as the observed TS-broken Chern insulators, imply an intriguing hybridization of conventional CDW and topological phase. Such combination of CDW and topological features is also numerically discovered in $\mathcal{C}=0,\pm1,2$ bands of several toy tight-binding models~\cite{PhysRevB.90.245106,CDW_Stefanos2,CDW_Stefanos4} and experimentally observed in other moir\'e materials~\cite{CI_Pierce,Polshyn2022}. The hybrid states in $|\mathcal{C}|=2$ bands may originate from ferromagnetism of internal degree of freedoms and their entanglement with real-space translation in high Chern number bands~\cite{PhysRevB.90.245106,Polshyn2022}. 

\subsection{Experiment in twisted bilayer graphene}
While the experiment above realizes the FCI states in the sense that they have no direct LL counterparts, a quite high magnetic field is still needed to produce the topological band structure. Therefore, to turn the dream of zero-field FCI into reality, we must find systems hosting intrinsic topological flat bands in the absence of (or at least with weak) external magnetic fields. As predicted by theoretical works~\cite{TBGCBand_Senthil,PhysRevLett.123.036401,PhysRevB.99.155415,PhysRevB.99.195455,TBGCBand_Senthil2}, TBG appears as a promising candidate to satisfy this requirement. A plethora of recent experiments down to zero field~\cite{Zondiner2020,CI_Yazdani,CI_Wu,CI_Young,Das2021,CI_Choi,CI_Park,CI_Stepanov,Serlin2020,CI_Pierce} have confirmed the existence of Chern insulating states at integer band fillings, thus indeed raising the possibility of realizing zero-field FCIs in TBG systems. 

\begin{figure}
	\centering
	\includegraphics[width=0.8\linewidth]{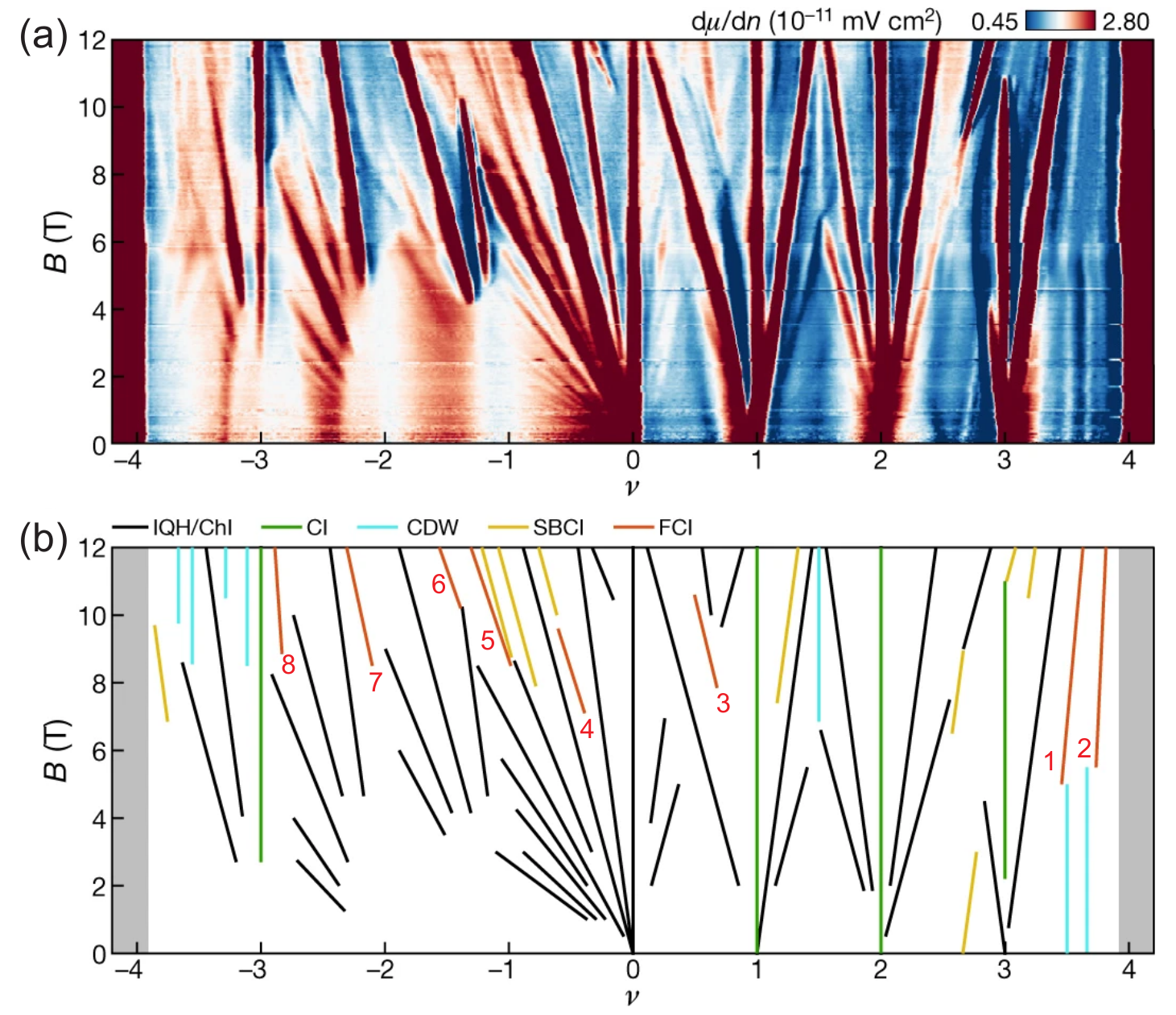}
	\caption{\label{fig:FCI_exp} Incompressible states observed in TBG-hBN. (a) Local inverse compressibility $d\mu/dn$ as a function of electron filling $\nu$ and the magnetic field $B$. Those red trajectories correspond to incompressible states. (b) Wannier diagram extracted from the experimental data in (a). Black lines correspond to integer quantum Hall states/TS-preserving Chern insulators (IQH/ChI); green lines correspond to trivial correlated insulators (CI); blue lines correspond to charge density waves (CDW); yellow lines correspond to TS-broken Chern insulators (SBCI); and orange lines correspond to (both TS-preserving and TS-broken) fractional Chern insulators (FCI). The detailed definitions of these states can be found in Table~\ref{tbl2}. The red numbers 1-8 indicate the eight observed FCI states. Figure modified from Ref.~\cite{FCIexp_Xie}. }
\end{figure}

A significant progress was made in this direction in which FCI states were observed in TBG aligned with hBN at weak magnetic fields as low as $\sim 5 \ {\rm T}$~\cite{FCIexp_Xie}. In this experiment, TBG devices were prepared near the magic twist angle $\theta\sim 1.06^\circ$, then a perpendicular magnetic field is applied. Local electronic compressibility measurements were preformed using a scanning single-electron transistor. Similar to the BLG-hBN experiment~\cite{FCI_Eric}, the dependence of this compressibility on the electron density and magnetic field, as shown in Fig.~\ref{fig:FCI_exp}(a), reveals a large number of incompressible states (Table~\ref{tbl2}) satisfying Eq.~(\ref{streda}), in which multiple FCI states with fractional $t$ and $s$ are identified. Remarkably, the first two FCI states appear in the range of $3<\nu<4$ at a weak magnetic field of only $\sim 5  \ {\rm T}$ [state 1 with $(t,s)=(2/3,10/3)$ and state 2 with $(t,s)=(1/3,11/3)$ in Fig.~\ref{fig:FCI_exp}(b)]. According to their $(t,s)$ values, these two states are formed at $\bar\nu=1/3$ and $\bar\nu=2/3$ filling of a $\mathcal{C}=-1$ band and preserve the TS of the moir\'e superlattice, thus they can be interpreted as lattice analogs of the $\nu=1/3$ Laughlin FQH state and its particle-hole conjugate. The energy gaps for both states are estimated as $\sim 0.6 \ {\rm K}$, which is comparable to the typical temperature required to realize Laughlin states in 2DEGs. Notably, these two FCI states are suddenly replaced by CDWs with the decreasing of the magnetic field. Numerical simulations in Refs.~\cite{FCIexp_Xie,FieldTunedFCI} interpreted this phenomenon in terms of quantum geometry of the partially filled band: the applied magnetic field flattens the band Berry curvature, such that favorable band geometry conditions for the emergence of FCIs are created. However, unlike in the BLG-hBN experiment where the whole Chern band is created by the magnetic field, here the Berry curvature stems from the zero-field topological band of TBG-hBN and the role of the magnetic field is merely to redistribute Berry curvature. Once the Berry curvature can be flattened by other effects, the magnetic field will probably become unnecessary and these $\bar\nu=1/3$ and $\bar\nu=2/3$ FCIs are expected to survive at zero field. Indeed, zero-field $\bar\nu=1/3$ FCIs in TBG-hBN were first numerically predicted at weaker $w_1=90 \ {\rm meV}$~\cite{FCITBG_Ahmed}, where the Berry curvature is flatter. Alternatively, a reduction of $w_0/w_1$ or slightly increasing the twist angle can also improve the band geometry~\cite{FieldTunedFCI}.

Apart from the weak-field FCI states 1 and 2, additional FCIs were also observed in TBG-hBN away from $3<\nu<4$ at slightly stronger magnetic fields, including TS-preserving states [state 6 with $(t,s)=(-13/5,-2/5)$ and state 7 with $(t,s)=(-4/3,-5/3)$] and TS-broken states [state 3 with $(t,s)=(-8/5,11/10)$ and state 4 with $(t,s)=(-7/3,2/9)$]. More intriguingly, there are also FCI states with coprime denominators of $t$ and $s$, like state 5 with $(t,s)=(-5/2,-1/5)$ and state 8 with $(t,s)=(-1/2,-8/3)$, which are neither simple TS-preserving nor TS-broken states. These exotic FCIs may originate from complex interplay between the multicomponent nature of TBG and spatial symmetry.

\section{FCIs in cold-atom systems}
Beside solid-state moir\'e systems, the high microscopic control and precision that are achievable in ultracold atoms make them also among the most promising candidates to realize rich correlated physics. In fact, there is a long-standing interest to realize the FQHE via cold atoms. The key task is to achieve a tight-binding model with complex hopping. Theoretical works along this direction have proposed several schemes, such as dipole systems~\cite{FCI_dipole,PhysRevLett.110.185302,Weber2022} and bosons in optical flux lattice~\cite{OFL_Cooper,OFL_Cooper2,FCI_OFL_Cooper,FCICA_Sterdyniak}, and predicted FCIs therein. In particular, tremendous effort is made to study onsite-interacting bosons confined in a square optical lattice penetrated by a uniform effective magnetic field. This system is described by the 2D Harper-Hofstadter-Hubbard (HHH) Hamiltonian
\begin{eqnarray} 
H=-J\sum_{m,n}\left(a^\dagger_{m,n+1} a_{m,n}+e^{i 2\pi \phi n} a^\dagger_{m+1,n} a_{m,n}+{\rm H.c.}\right)+
\frac{U}{2}\sum_{m,n}\hat{n}_{m,n}\left(\hat{n}_{m,n}-1\right),
\label{HHH}
\end{eqnarray} 
where $J$ is the nearest-neighboring hopping strength, $a^\dagger_{m,n}$ ($a_{m,n}$) creates (annihilates) a boson on the lattice site $(m,n)$, $\phi$ is the effective magnetic flux through each elementary square of the lattice, $U$ is the magnitude of onsite repulsion, and $\hat{n}_{m,n}=a^\dagger_{m,n}a_{m,n}$ is the occupation number operator on site $(m,n)$.  A plethora of numerical works find bosonic Abelian and non-Abelian FCIs as ground states of the HHH model at rational flux density $\phi\leq 1/3$ when the interaction is sufficiently strong and the lowest band is partially filled~\cite{PhysRevLett.96.180407,FQH_OL_Anders,FCICA_Lukin,FQH_OL_Simon,PhysRevB.96.195123,GunnarHof,PhysRevA.100.053624,PhysRevB.103.L161101,PhysRevB.105.075135}. The introduction of longer-range hopping and reduction of flux density may further stabilize these states. Fine-tuned exponentially decaying hopping can even perfectly flatten the lowest band and support model FCIs as the zero-energy ground states~\cite{KapitMueller,ZhaoKapit}. 

The FCI states found in the HHH model can be prepared quasi-adiabatically, that is, one begins with a trivial state with low entropy, then slowly ramps the system to the desired final state~\cite{PhysRevA.70.053612,PhysRevLett.113.155301,FCI_dipole2,PhysRevB.96.201103,FCI_OL_Pollmann,PhysRevB.103.L161101}. This protocol replies on continuous phase transitions between FCIs and other competing phases like CDW and superfluid. While dissipation is often thought of being harmful to the success of state preparation, some protocols suggest using special dissipative mechanism to pumping bosons from higher bands to the lowest band to make the state closer to FCIs~\cite{PhysRevResearch.3.043119}.

Theoretical works are further encouraged by the rapid development of experimental techniques on artificial gauge field and quantum gas microscope, which make the HHH Hamiltonian a paradigmatic model that has been realized in cold-atom experiments. Once bosons are trapped in an optical lattice, the effective magnetic field can be achieved by generating an artificial gauge field through lattice shaking techniques~\cite{Hofstadter_Bloch,Hofstatder_Miyake}. The nontrivial band topology in the HHH model has been confirmed by measuring the transverse deflection of the atomic cloud in response to an optical gradient~\cite{Aidelsburger2015}. Further, the model has been pushed into the strong-interaction regime~\cite{Tai2017}, such that FCIs of ultracold atomic gases may potentially be within reach in the not too distant future.

Motivated by the encouraging theoretical and experimental progress, recent research focuses on how to detect FCI states once they are ready in cold-atom systems. In particular, such detection should be experimentally feasible  and valid for those few-particle settings limited by the current technology. One way of detection is to probe the Hall conductance. However, because the transport measurements akin to those in solid state systems cannot be straightforwardly adopted into the cold-atom setting, alternative schemes have to be designed to extract Hall conductance. There have been several protocols on this aspect. One can apply a weak external force to the atomic cloud and track its center of mass drift~\cite{PhysRevA.102.063316,FCICA_Motruk}. It is also possible to evaluate Hall conductance using interferometric protocols~\cite{Grusdt2016}, randomized measurement scheme~\cite{MBChern_Hafezi}, or through circular-dichroic measurement~\cite{Circular_Goldman,Circular_Ceceil}. Apart from Hall conductance measurements, other promising schemes include probing fractionalized quasiholes through density measurement~\cite{FCIHole_Eckardt,PhysRevA.98.063629,FCIHole_Macaluso,FCIHole_Eckardt2} and detecting chiral edge excitation~\cite{PhysRevB.85.235137,ChiralEdge_Frank}. All of these proposals have been examined by numerical simulations in small cold-atom systems. Yet, it remains unclear which protocol is the most experimentally feasible before FCIs of cold atoms really turn to reality.

\section{Conclusions and outlook}
In this chapter, we have presented an introduction to fractional Chern insulators in topological flat bands. While FCIs share key features with conventional FQH states, the significant discrepancies between a Chern band and a Landau level make FCIs an ideal stage to realize the FQH physics under more general experimental conditions, to explore in the pursuit for qualitatively new topologically ordered states, and for unveiling the profound interplay between interaction, topology and geometry. Although it is impossible to provide a fully comprehensive account of the entire field here, we hope that our introduction has demonstrated the major methodology that has been used to understand FCIs and the new research trends which are emerging recently in this field. 

The enthusiasm of studying FCIs has lasted over the past decade, and is now gaining renewed and extended interest due to their direct relevance in moir\'e materials. We would like to close this chapter by listing several directions which in our opinion are important for the development of this field in near future. 

\begin{itemize}

\item
{\bf Exploring new platforms.}
The recent trend of studying FCIs in moir\'e materials has clearly demonstrated the intimate relation between FCI and material science. Searching for new solid-state platforms that can host FCIs will significantly stimulate further development of this field. The research in this direction will need contributions from material calculations via, for instance, density functional theory. In fact, progress is already being made in this direction. In addition to more graphene based materials~\cite{tMBG_Park,tMBG_Louk,tMBG_Chen}, moir\'e systems built on transition metal dichalcogenides (TMD) are found to support topological flat bands~\cite{TMD_Wu2019,TMD_DasSarma2020,Li2021,TMD_Fu2021,TMD_Fu2021_2,TMD_DasSarma2021,TMD_Zhou2022}. Excitingly, a few numerical works have identified FCIs in twisted TMD homobilayers, like twisted bilayer ${\rm MoTe}_2$~\cite{TMDFCI_Li2021} and ${\rm WSe}_2$~\cite{FCITMD_Fu}. It requires more effort to thoroughly investigate FCI states and their competing phases in new materials. These theoretical explorations will be greatly helpful to accelerate the experimental realization of zero-field FCIs.

\item
{\bf Experimental realization of high-temperature zero-field FCIs.}
Despite of the encouraging experimental observation of FCI states in TBG-hBN at weak magnetic fields down to $\sim5 \ {\rm T}$~\cite{FCIexp_Xie}, how to stabilize these states at further reduced fields remains a crucial challenge for really achieving zero-field FCIs. Theoretical works have suggested to probe TBG-hBN at twist angles above the magic angle, which could improve the band geometry such that the magnetic field might become unnecessary~\cite{FieldTunedFCI}. Alternatively, one can also try other materials like twisted bilayer TMD. These proposals should be examined by experiments, then the optimal setups which facilitate the realization of zero-field FCIs the most can be selected. Moreover, apart from the compressibility measurement used in current experiments, transport features, which directly show the characterizing Hall conductance plateaus, is in an ideal situation a preferable method to identify FCIs in the future.

In addition to the static realization of zero-field FCIs, one may also consider the dynamic preparation of these states in periodically driven systems. Recently, there have been a series of works studying the photon-dressed band structure of moir\'e materials driven by an external laser light field~\cite{Sentef2019,Katz2020,Vogl2020,Yun2020,Rodriguez2020,Assi2021,Topp2021,Vogl2021,Benlakhouy2022}. Based on these Floquet band structures, numerical simulations have demonstrated the possibility of Floquet FCIs~\cite{Grushin2014,Anisimovas2015} in light-driven TBG~\cite{FloquetFCI_Zhao}. 

The energy gaps of the observed $\bar\nu=1/3$ FCIs in TBG-hBN are about $0.6 \ {\rm K}$~\cite{FCIexp_Xie}, which are still close to that of the conventional Laughlin state in 2DEGs. Considering the expectation that FCIs can in principle exist at much higher temperature than conventional FQH states, it is definitely worthy of seeking a way to further increase the gap.

\item
{\bf  Searching and understanding new states.}
With more theoretical and experimental explorations in various platforms, we should pay special attention on searching and understanding new states. Clearly, non-Abelian FCI states are still absent in experimental observation. In fact, it is difficult to get them even in numerical simulations, where multi-body interactions are often required~\cite{FCIChern_Wang2,FCIsymmetry,FCIZoology,FCIHighC_Sterdyniak,ModelFCI_Zhao}. Considering the potential application of non-Abelian anyons in topological quantum computation, it is a challenging albeit urgent task to find schemes that can provide non-Abelian FCIs by realistic two-body interactions. 

The valley and spin degrees of freedom in moir\'e systems make it interesting to probe novel multicomponent FCI states. On the other hand, the symmetry-broken and other exotic FCI states (Table~\ref{tbl2}) found in the recent TBG-hBN experiment~\cite{FCIexp_Xie} fall beyond the scope of conventional translational-symmetry-preserving FCIs. Apparently they have no analog in continuum FQH states. More theoretical and experimental effort is needed to understand these enigmatic states. Crucial open questions include: (i) Can these states survive at zero magnetic field? (ii) Can numerical simulations probe these states and extract their topological orders? (iii) Can experiments further identify the nature of these states, for example, their quasihole charges?    

\end{itemize}

Motivated by seeking solutions of the questions above, we are convinced that the study of FCIs will remain an active field providing exciting and beautiful new physics.

\section{Acknowledgements}
Z. L. is supported by the National Key Research and Development Program of China through Grant No. 2020YFA0309200.

\bibliographystyle{plainnat}
\bibliography{mybibfile}

\end{document}